\documentclass[reprint,aps,pra,showpacs,floatfix,twocolumn,nofootinbib]{revtex4-1}
\usepackage{graphicx}
\usepackage{dsfont}
\usepackage{epstopdf}
\usepackage{amsfonts}
\usepackage{amsmath}
\usepackage[toc, page]{appendix}
\usepackage{multirow}

\usepackage{color}

\newcommand{\be}{\begin{equation}}
\newcommand{\ee}{\end{equation}}

\begin{document}

\title{Suitable bases for quantum walks with Wigner coins}

\author{I. Bezd\v ekov\'a}
\affiliation{Department of Physics, Faculty of Nuclear Sciences and Physical Engineering, Czech Technical University in Prague, B\v
rehov\'a 7, 115 19 Praha 1 - Star\'e M\v{e}sto, Czech Republic}

\author{M. \v Stefa\v n\'ak}
\affiliation{Department of Physics, Faculty of Nuclear Sciences and Physical Engineering, Czech Technical University in Prague, B\v
rehov\'a 7, 115 19 Praha 1 - Star\'e M\v{e}sto, Czech Republic}

\author{I. Jex}
\affiliation{Department of Physics, Faculty of Nuclear Sciences and Physical Engineering, Czech Technical University in Prague, B\v
rehov\'a 7, 115 19 Praha 1 - Star\'e M\v{e}sto, Czech Republic}

\date{\today}

\begin{abstract}
The analysis of a physical problem simplifies considerably when one uses a suitable coordinate system. We apply this approach to the discrete-time quantum walks with coins given by $2j+1$-dimensional Wigner rotation matrices (Wigner walks), a model which was introduced in T. Miyazaki et al., Phys. Rev. A \textbf{76}, 012332 (2007). First, we show that from the three parameters of the coin operator only one is physically relevant for the limit density of the Wigner walk. Next, we construct a suitable basis of the coin space in which the limit density of the Wigner walk acquires a much simpler form. This allows us to identify various dynamical regimes which are otherwise hidden in the standard basis description. As an example, we show that it is possible to find an initial state which reduces the number of peaks in the probability distribution from generic $2j+1$ to a single one. Moreover, the models with integer $j$ lead to the trapping effect. The derived formula for the trapping probability reveals that it can be highly asymmetric and it deviates from purely exponential decay. Explicit results are given up to the dimension five.
\end{abstract}

\maketitle

\section{Introduction}

Quantum walk \cite{adz,meyer,fg} is an extension of the concept of a random walk \cite{hughes} to unitary evolution of a quantum particle on a graph. Both continuous-time and discrete-time (or coined) quantum walks have found promising applications in quantum information processing, e.g. in quantum search algorithms \cite{skw,childs:search:2004,childs:search:2014,vasek:search}, graph isomorphism testing \cite{gamble,berry,rudiger}, finding structural anomalies in graphs \cite{reitzner,hillery,cottrell} or as universal tools for quantum computation \cite{childs,Lovett}. Quantum walks were realized in a number of experiments using optically trapped atoms \cite{karski}, cold ions \cite{schmitz,zahringer} and photons \cite{and:1d,broome,peruzzo:corelated:photons,two:photon:waveguide,sansoni,and:2dwalk:science,delayed:choice}.

The dynamics of the discrete-time quantum walks crucially depends on the choice of the coin operator and the initial state \cite{tregenna}, which has essential implications for the performance of the search algorithm \cite{ambainis:search}. The behaviour of two-state quantum walks with homogeneous coin is fully understood \cite{konno:limit:2002}. This is not true for quantum walks with higher-dimensional coins, since the complexity of the unitary group grows rapidly with its dimension. Most of the known results were obtained for quantum walks with Grover coin which leads to the so-called trapping (or localization) effect \cite{inui:grover1,inui:grover2}. In contrast to the two-state walk, which show purely ballistic spreading, the particle performing the Grover walk has a non-vanishing probability to remain trapped in the vicinity of the origin. This is represented by a stationary peak in the probability distribution located at the origin, which does not vanish with the increasing number of steps but decays exponentially with the distance from the starting point. The Grover walk and its extensions were intensively studied, either for line \cite{inui:grover1,inui:grover2,stef:cont:def,falkner,stef:limit,machida,stef:3state}, plane \cite{inui:2dgrover,watabe:grover,balint:strong:trap} or higher dimensional lattices \cite{hinajeros:grover,higuchi:grover}.

Among the few models of quantum walks on a line which were solved analytically for arbitrary dimension of the coin operator, i.e. arbitrary number of displacements particle can make in a single time-step, are the Wigner walks introduced in \cite{konno:wigner}. This model utilizes $2j+1$-dimensional Wigner rotation matrices as coin operators and is closely related to the tensor product model of quantum walks \cite{mackay:2002,brun:2003:pra,brun:2003:prl,andraca:2005}. Explicitly, using the Fourier analysis \cite{ambainis} and the weak-limit theorem \cite{Grimmett}, the limit density of the Wigner walks was derived in \cite{konno:wigner} for arbitrary $j$. However, as the authors point out in \cite{konno:wigner}, the dependence of the limit density on the initial coin state is rather involved. The reason for this inconvenience is that the initial coin state is decomposed into the standard basis of the coin space, i.e. the one in which the step operator is defined. The aim of our paper is to present an alternative approach to \cite{konno:wigner}. Namely, by using a more suitable basis of the coin space we put the results of \cite{konno:wigner} into much more convenient form. This allows us to identify previously unknown features which are hidden in the standard basis description. As an example, we show that while the generic probability distribution of the $2j+1$-dimensional Wigner walk has $2j+1$ peaks, there exist initial conditions for which the number of peaks reduces to one. Moreover, the models with integer $j$ exhibit the trapping effect similar to the Grover walk. This feature was not analyzed in detail in \cite{konno:wigner}. We derive the explicit formula for the trapping probability for $j=1$ and $j=2$. Our results show that the trapping probability can be highly asymmetric. Moreover, the decay of the trapping probability with the distance from the origin is not exactly exponential for $j=2$.

The paper is organized as follows. In Section~\ref{review} we review the results on quantum walks with Wigner coin obtained in \cite{konno:wigner}. We find that when we focus only on the limit density and its moments the number of physically relevant coin parameters can be reduced from three to one. Next, we outline the procedure to determine the suitable basis of the coin space in which the limit density simplifies considerably. Namely, we choose the basis states among those which give rise to non-generic distributions. In Section~\ref{sec:1/2} the suitable basis is determined for the simplest case of the two-state model. Moreover, we find a recipe for the construction of suitable bases for higher-dimensional models. We apply this recipe in Section~\ref{sec:1} to the three-state model and we find that it is equivalent to the three-state Grover walk \cite{inui:grover1,inui:grover2} and its one-parameter generalization \cite{stef:cont:def,falkner,stef:limit,machida}. In Sections~\ref{sec:3/2} and \ref{sec:2} the four-state and five-state models are studied, respectively. Expressing the limit density in the suitable basis allows us to determine additional non-generic situations. In particular, we find initial conditions such that the number of peaks in the probability distribution is reduced to one. Finally, we summarize our results and present an outlook in Section~\ref{sec:outlook}. More technical details of the analysis are left for the Appendices.

\section{Quantum walks with Wigner coins}
\label{review}

In this Section we review the quantum walks on a line with Wigner coins (Wigner walks) following their introduction and analysis in \cite{konno:wigner}. The Hilbert space of the Wigner walk is given by a tensor product
$$
{\cal H} = {\cal H}_P\otimes{\cal H}_C,
$$
of the position space
$$
{\cal H}_P = {\rm Span}\left\{|x\rangle|x\in\mathds{Z}\right\},
$$
and the coin space ${\cal H}_C$. The dimension of the coin space is $2j+1$, where $j$ is a (half-)integer. The standard basis of the coin space is formed by vectors $|m\rangle$ corresponding to the jumps of length $2m$ where $m=-j,-j+1 \dots, j$. Single step of the time-evolution is given by a unitary operator
\begin{equation}
\label{evol:U}
\hat{U} = \hat{S}\cdot(\hat{I}\otimes \hat{R}^{(j)}(\alpha, \beta, \gamma)),
\end{equation}
where $\hat{S}$ is the step operator
\begin{equation}
\label{step}
\hat{S} = \sum_{x=-\infty}^{+\infty} \sum_{m=-j}^j |x+2m\rangle\langle x|\otimes |m\rangle\langle m|,
\end{equation}
and $\hat{R}^{(j)}(\alpha, \beta, \gamma)$ denotes the coin operator which is given by the Wigner rotation matrix \cite{31,32}, i.e. the $2j+1$-dimensional irreducible representation of the rotation group SO(3). The matrix elements of the coin in the standard basis of the coin space
$$
R^{(j)}_{mn}(\alpha, \beta, \gamma)=\langle m|\hat{R}^{(j)}(\alpha, \beta, \gamma)|n\rangle,
$$
are given by
$$
R^{(j)}_{mn}(\alpha, \beta, \gamma)=e^{-i \alpha m}r^{(j)}_{mn}(\beta)e^{-i\gamma n},
$$
where
\begin{widetext}
$$
r^{(j)}_{mn}(\beta) = \sum_l \Gamma(j,m,n,l) \left(\cos\frac{\beta}{2}\right)^{2j+m-n-2l}\left(\sin\frac{\beta}{2}\right)^{2l-m+n}.
$$
The factor $\Gamma(j,m,n,l)$ reads
\begin{equation}
\label{Gamma}
\Gamma(j,m,n,l) = (-1)^l \frac{\sqrt{(j+m)!(j-m)!(j+n)!(j-n)!}}{(j-n-l)!(j+m-l)!(l-m+n)!l!}.
\end{equation}
\end{widetext}
The summation index $l$ runs over all integers such that all factorials in (\ref{Gamma}) are well defined.

As the initial state of the Wigner walk we consider
$$
|\psi(0)\rangle = |0\rangle\otimes|\psi_C\rangle,
$$
i.e. the particle starts from the origin with the initial coin state
\begin{equation}
\label{initC}
|\psi_C\rangle = \sum\limits_{m=-j}^j q_m|m\rangle.
\end{equation}
The amplitudes $q_m$ fulfill the normalization condition
$$
\sum\limits_{m=-j}^j |q_m|^2 = 1.
$$
The state of the particle after $t$ steps of the walk reads
$$
|\psi(t)\rangle = \hat{U}^t|\psi(0)\rangle,
$$
which can be decomposed into the standard basis as
$$
|\psi(t)\rangle = \sum_x\sum_{m=-j}^j \Psi_m^{(j)}(x,t)|x\rangle\otimes|m\rangle.
$$
The probability to find the particle at position $x$ after $t$ steps of the quantum walk is then given by
$$
p^{(j)}(x,t) = \sum_{m=-j}^j \left|\Psi_m^{(j)}(x,t)\right|^2.
$$
We note that all Wigner walks are bipartite, i.e. half of the lattice points are empty at any time step. For integer $j$ the odd sites are never occupied, while for half-integer $j$ the walk oscillates between even and odd sites.

Since the Wigner walks are translationally invariant, their analysis is greatly simplified in the Fourier picture \cite{ambainis}. Moreover, in the asymptotic limit the moments of the particle's re-scaled position (or pseudo-velocity) can be expressed in the form \cite{Grimmett}
\begin{equation}
\label{moment}
\lim\limits_{t\rightarrow +\infty} \left\langle\left(\frac{x}{t}\right)^n\right\rangle = \int v^n \nu^{(j)}(v) dv,
\end{equation}
where $\nu^{(j)}(v)$ is the limit density. Its explicit form for Wigner walks was derived in \cite{konno:wigner} where it was shown that it is given by a sum
\begin{equation}
\label{totaldens}
\nu^{(j)}(v) = \sum_{0<m \leq j} \nu^{(j,m)}(v).
\end{equation}
The summation index $m$ runs over (half-)integers, depending on whether $j$ is half-integer or integer, in unit steps. The individual densities $\nu^{(j,m)}(v)$ have the form
\begin{equation}
\label{indiv:dens}
\nu^{(j,m)}(v) = \frac{1}{2m} \mu\left(\frac{v}{2m};\cos\frac{\beta}{2}\right) {\cal M}^{(j,m)}\left(\frac{v}{2m}\right)
\end{equation}
where $\mu(v;a)$ is the Konno's density function \cite{Konno:2002a,Konno:2005}
\begin{equation}
\label{mu}
\mu(v;a) = \frac{\sqrt{1-a^2}}{\pi(1-v^2)\sqrt{(a-v)(a+v)}}.
\end{equation}
The symbol ${\cal M}^{(j,m)}(v)$ denotes the weight function which is a polynomial of degree $2j$ in $v$ with coefficients determined by the initial state and the coin parameters $\beta$ and $\gamma$. The decomposition (\ref{totaldens}) of the limit density $\nu^{(j)}(v)$ into $\lceil j\rceil$ densities $\nu^{(j,m)}(v)$ indicates that the $2j+1$-state Wigner walk can be considered as superposition of $\lceil j\rceil$ walks which propagates through the lattice with different velocity given by $2m\cos\frac{\beta}{2}$. In addition to the evaluation of moments (\ref{moment}), the limit density (\ref{totaldens}) can be used to approximate the shape of the probability distribution of the Wigner walk in finite time according to
\begin{equation}
\label{dist:approx}
p^{(j)}(x,t) \approx \frac{2}{t}\nu^{(j)}(\frac{x}{t}).
\end{equation}
The factor of 2 accounts for the fact that the Wigner walks are bipartite. Note that the Konno's density function $\mu(v;a)$ diverges for $v\rightarrow\pm a$. These divergencies correspond to the $2j+1$ peaks ($2j$ if $j$ is an integer) in the probability distribution $p^{(j)}(x,t)$ which are found at $x\approx \pm  2m\cos\frac{\beta}{2} t$, where the range of the index $m$ is the same as in the sum (\ref{totaldens}).

We note that for integer $j$ the particle is allowed to stay at its actual position, while for half-integer $j$ it has to leave the previously occupied site. This has a crucial impact on the dynamics of the quantum walk, as was found in \cite{konno:wigner}. Indeed, for half-integer $j$ the spectrum of the evolution operator (\ref{evol:U}) is purely continuous, while for integer $j$ it has, in addition, an isolated eigenvalue with infinitely-many localized eigenstates. The presence of the point spectrum results in an additional peak in the center of the probability distribution which decays rapidly with the distance from the origin. Moreover, the peak does not vanish with increasing number of time steps, i.e. for integer $j$ the particle has a non-vanishing probability to remain at position $x$ in the asymptotic limit. This feature was first observed for three-state Grover walk on a line \cite{inui:grover1,inui:grover2}. We denote the limiting value of the probability to remain at position $x$ as
\begin{equation}
\label{trap:prob}
p^{(j)}_\infty(x)=\lim\limits_{t\rightarrow\infty}p^{(j)}(x,t),
\end{equation}
and call it the trapping probability. Its explicit form was not given in \cite{konno:wigner}. We evaluate the trapping probability in Sections~\ref{sec:1} and \ref{sec:2}, where we treat the Wigner walks with $j=1$, resp. $j=2$, following the approach used in \cite{falkner,stef:limit,machida}.

In principle, the formulas (\ref{totaldens}), (\ref{indiv:dens}) and the general expression for the weight functions ${\cal M}^{(j,m)}(v)$ derived in \cite{konno:wigner} allow us to calculate the limit density $\nu^{(j)}(v)$ for arbitrary (half-)integer $j$. However, the actual form of the limit density is rather complicated already for small values of $j$, as can be observed in \cite{konno:wigner} where the authors provided the explicit results up to $j=\frac{3}{2}$. This makes further investigation of Wigner walks quite demanding. It is the aim of the paper to simplify the limit density as much as possible. In order to do so, we first discuss which of the coin parameters are physically relevant. Second, we simplify the dependence of the weight function on the initial coin state by choosing a more suitable basis in the coin space. Since we are interested only in the moments and the limit density, we consider two Wigner walks with coins $\hat{R}^{(j)}(\alpha_1,\beta_1,\gamma_1)$ and $\hat{R}^{(j)}(\alpha_2,\beta_2,\gamma_2)$ as equivalent if for any initial coin state $|\psi_{C_1}\rangle$ of the first walk there exists an initial coin state of the second walk $|\psi_{C_2}\rangle$ such that the resulting limit densities are the same. We note that two equivalent Wigner walks, as we have defined them, might exhibit different properties when additional aspects of quantum walks, such as topological phases \cite{tp:kitagawa,tp:obuse,tp:kitagawa2,tp:asboth,tp:kitagawa3,tp:asboth2,tp:moulieras,tp:tarasinski}, are of interest.

Let us begin with the coin parameters. Notice that Konno's density function depends only on one of them, namely $\beta$. Moreover, as was shown in \cite{konno:wigner} the weight functions ${\cal M}^{(j,m)}$ are determined by the initial coin state and the coin parameters $\beta$ and $\gamma$, but they are independent of $\alpha$. Hence, all quantum walks with different $\alpha$ are equivalent, as far as the limit density and the moments of the particle's position are concerned. Therefore, we consider $\alpha=0$ from now on. Moreover, the dependence of the weights ${\cal M}^{(j,m)}$ on $\gamma$ is rather simple. As was found in \cite{konno:wigner} the parameter $\gamma$ enters only through the terms of the form $q_m \overline{q_n} e^{-i(m-n)\gamma}$. Hence, the Wigner walk with coin $\hat{R}^{(j)}(0,\beta,0)$ and the initial coin state (\ref{initC}) gives the same limit density as the Wigner walk with the coin $\hat{R}^{(j)}(0,\beta,\gamma)$ and the initial coin state
$$
|\psi_C^\gamma\rangle = \sum\limits_{m=-j}^j q_m e^{im\gamma}|m\rangle.
$$
Therefore, all models with different $\gamma$ are equivalent according to our definition and we restrict our further analysis to the case $\gamma=0$. Hence, we are left with only one physically relevant coin parameter $\beta$. We note that for two-state walks the equivalence of the three-parameter set of quantum walks with coins ${\hat R}^{(1/2)}(\alpha,\beta,\gamma)$, which in fact covers all translationally invariant two-state quantum walks, with the single-parameter family ${\hat R}^{(1/2)}(0,\beta,0)$ was already established in \cite{unitary:equivalence}. The above discussion shows that within the set of Wigner walks the equivalence hold for arbitrary $j$.

We now turn to a slightly different parametrization of the coin operator ${\hat R}^{(j)}(0,\beta,0)$ which is more suitable in the context of Wigner walks. Namely, instead of $\beta$ we consider the parameter $\rho$ given by
$$
\rho = \cos{\frac{\beta}{2}},
$$
which corresponds to the divergencies of the limit densities (\ref{indiv:dens}). It also directly determines the speed of propagation of the wave-packet through the lattice \cite{kempf}. Since the Euler angle $\beta$ is limited to the interval $[0,\pi]$, the new parameter $\rho$ varies from 0 to 1, and the identity
$$
\sin{\frac{\beta}{2}} = \sqrt{1-\rho^2},
$$
holds. Finally, we define the coin operator ${\hat R}^{(j)}(\rho) \equiv {\hat R}^{(j)}(0,\beta,0)$ with the matrix elements in the standard basis given by
\begin{eqnarray}
\label{matrix:rho}
\nonumber R^{(j)}_{mn}(\rho) & = & \langle m|{\hat R}^{(j)}(\rho)| n\rangle = \sum_l \Gamma(j,m,n,l) \\
& & \times \rho^{2j+m-n-2l}\left(\sqrt{1-\rho^2}\right)^{2l-m+n}.
\end{eqnarray}

Let us now turn to the dependence of the limit density on the initial coin state.
As was pointed out in \cite{konno:wigner}, the weight functions ${\cal M}^{(j,m)}$ are rather involved functions of the coefficients $q_i$ of the initial coin state in the standard basis. The aim of this paper is to simplify these expressions by choosing a more suitable basis of the coin space. Following the common experience of quantum mechanics one would expect that the suitable basis is the one given by the eigenvectors of the operator involved. This is indeed the case of the three-state Grover walk \cite{stef:limit}. However, for Wigner walks the eigenvectors do not represent the best choice. To construct the suitable basis we consider one additional property of the eigenvectors of the Grover coin used in \cite{stef:limit}. Namely, the basis vectors were chosen such that they result in non-generic distributions of the Grover walk, i.e. they reduce the number of peaks in the probability distribution. We adopt this requirement for Wigner walks and select the suitable basis vectors among such states. The conditions for non-generic distributions are straightforward in the asymptotic limit, since then the peaks correspond to the divergencies of the limit density (\ref{totaldens}). Therefore, we have to determine the states for which some of the divergencies of (\ref{totaldens}) vanish. In the following Section we solve these conditions and determine the suitable basis states for the two-state model. We then rewrite the suitable basis in terms of the eigenvectors of the coin operator. This gives us a recipe for construction of suitable bases in higher-dimensional models.

\section{Two-state model}
\label{sec:1/2}

We begin with the simplest case of a two-state model where $j=\frac{1}{2}$. The coin space is two-dimensional with the basis states  $|1/2\rangle,\ |-1/2\rangle$, which correspond to the jumps of one step to the right and one step to the left. In the standard basis the matrix representation of the coin operator is given by
\begin{equation}
\label{coin1/2}
R^{(1/2)}(\rho) =\left(
                   \begin{array}{cc}
                    \rho & -\sqrt{1-\rho^2} \\
                     \sqrt{1-\rho^2} & \rho \\
                   \end{array}
                 \right).
\end{equation}
The limiting probability density calculated in \cite{konno:wigner} reads
\begin{equation}
\label{nu}
	\nu^{(1/2)}(v) = \mu(v;\rho){\cal M}^{(1/2,1/2)},
\end{equation}
where $\mu(v;\rho)$ is the Konno's density function (\ref{mu}) and the weight ${\cal M}^{(1/2,1/2)}$ is given by
\begin{equation}
\label{weight:1/2}
{\cal M}^{(1/2,1/2)} = 1 + {\cal M}_1^{(1/2,1/2)}v.
\end{equation}
The linear term ${\cal M}_1^{(1/2,1/2)}$ in the weight has the form
\begin{eqnarray}
\label{weight:1/2:1}
\nonumber {\cal M}_1^{(1/2,1/2)} & = & -|q_{1/2}|^2 + |q_{-1/2}|^2 + \\
 & & +\frac{\sqrt{1-\rho^2}}{\rho}(q_{1/2}\overline{q_{-1/2}} + \overline{q_{1/2}}q_{-1/2}),
\end{eqnarray}
where $q_i$'s represent the coefficients of the initial coin state in the standard basis of the coin space
\begin{equation}
\label{init:2state}
|\psi_C\rangle = q_{1/2}|1/2\rangle + q_{-1/2}|-1/2\rangle.
\end{equation}
We will now determine the suitable basis for the description of the two-state model. Following the discussion in the previous section, we will construct the suitable basis from the states which lead to non-generic probability distributions, i.e. the states for which one of the divergencies of the limit density (\ref{nu}) vanishes. These divergencies coincide with those of the Konno`s density function (\ref{mu}), which appear for $v = \pm \rho$. To eliminate them we have to find such $q_{\pm 1/2}$ that the weight function tends to zero faster than the denominator of the Konno's density function for $v = \pm \rho$, i.e. the weight has to attain the form
$$
{\cal M}^{(\frac{1}{2}, \frac{1}{2})}= 1 \pm\frac{v}{\rho}.
$$
Hence, the linear term (\ref{weight:1/2:1}) has to be equal to $\pm \frac{1}{\rho}$. The solutions of these two equations
\begin{eqnarray*}
1-\frac{v}{\rho} :& &\;\;\;\;q_{ 1/2} =  \sqrt{\frac{1 + \rho}{2}}\nonumber\\
	     & &\;\;\;q_{- 1/2} = - \sqrt{\frac{1 - \rho}{2}},\nonumber\\
1+\frac{v}{\rho} :& &\;\;\;q_{ 1/2} = \sqrt{\frac{1 - \rho}{2}},\nonumber\\
         & & \;\;\;q_{- 1/2} = \sqrt{\frac{1 + \rho}{2}},
\end{eqnarray*}
provide the coefficients of the initial states (\ref{init:2state}) that eliminate the divergence of the limit density (\ref{nu}) at $v=\pm\rho$. We denote these vectors as $|\chi^\pm\rangle$. In the standard basis they have the form
\begin{eqnarray}
\label{basis:1/2:st}
\nonumber |\chi^+\rangle & = & \sqrt{\frac{1+\rho}{2}}|1/2\rangle - \sqrt{\frac{1-\rho}{2}}|-1/2\rangle,\\
|\chi^-\rangle & = & \sqrt{\frac{1-\rho}{2}}|1/2\rangle + \sqrt{\frac{1+\rho}{2}}|-1/2\rangle.
\end{eqnarray}
Clearly, these two states form an orthonormal basis of the coin space. Moreover, we find that in this basis the weight function (\ref{weight:1/2}) simplifies considerably. We express the initial coin state in terms of the basis $\left\{|\chi^\pm\rangle\right\}$ according to
$$
|\psi_C\rangle = h^+|\chi^+\rangle + h^-|\chi^-\rangle.
$$
The correspondence between the amplitudes of the initial coin state in the suitable basis $h_i$ and in the standard basis is then given by
\begin{eqnarray}
\nonumber q_{1/2} & = & \sqrt{\frac{1+\rho}{2}} h^+ + \sqrt{\frac{1-\rho}{2}} h^-, \\
\nonumber q_{-1/2} & = & -\sqrt{\frac{1-\rho}{2}} h^+ + \sqrt{\frac{1+\rho}{2}} h^-.
\end{eqnarray}
Inserting these relations into the formula (\ref{weight:1/2:1}) we find that the linear term of the weight function reduces into
$$
{\cal M}_1^{(1/2,1/2)} = \frac{1}{\rho}(1-2|h^+|^2).
$$
Hence, the limit density for a two-state quantum walk with Wigner coin (\ref{coin1/2}) in the suitable basis $\left\{|\chi^\pm\rangle\right\}$ reads
\begin{equation}
\label{limit:dens:1/2}
\nu^{(1/2)}(v) = \frac{\sqrt{1-\rho^2}\left(1+(1-2|h^+|^2)\frac{v}{\rho}\right)}{\pi (1-v^2)\sqrt{(\rho-v)(\rho+v)}}.
\end{equation}
This result shows another benefit of using the basis $\left\{|\chi^\pm\rangle\right\}$. Namely, by absorbing part of the $\rho$ dependence into the definition of the basis states  (\ref{basis:1/2:st}), $\rho$ is effectively reduced to a scaling parameter. Indeed, varying the value of $\rho$ while keeping the amplitudes $h^+$ and $h^-$ untouched does not affect the shape of the limit density (\ref{limit:dens:1/2}). The parameter $\rho$ simply determines how far the density is stretched. We point out that the same will apply for Wigner walks with higher values of $j$.

\begin{figure}[htbp]
\begin{center}
\includegraphics[width=0.45\textwidth]{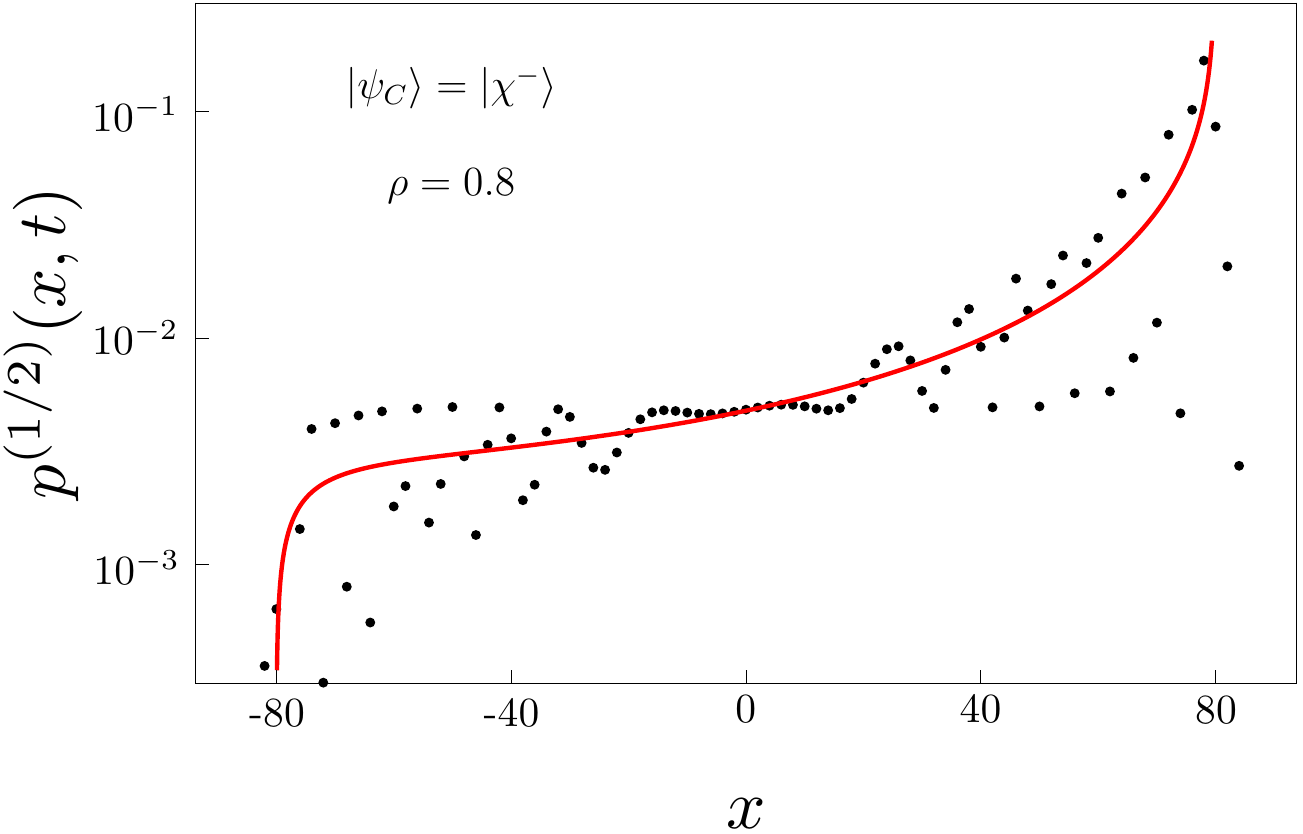}
\end{center}
\caption{(Color
  online) Probability distribution after 100 steps of the two-state Wigner walk with the coin $R^{(1/2)}(\rho)$. The initial state was chosen as $|\chi^-\rangle$. The coin parameter is $\rho=0.8$. We note that for different values of $\rho$ the probability distribution will have the same shape, only its width will change accordingly. The same applies to all figures.}
\label{fig:1}
\end{figure}

To illustrate our results we display in FIG.~\ref{fig:1} the probability distribution for the initial coin state $|\chi^-\rangle$ after 100 steps. The red curve depicts the limit density which for $|\chi^-\rangle$ is given by
$$
\nu^{(1/2)}_{|\chi^-\rangle} (v) = \frac{\sqrt{1-\rho^2}\sqrt{\rho+v}}{\pi \rho(1-v^2)\sqrt{\rho-v}}.
$$
Clearly, the limit density diverges for $v\rightarrow\rho$ but tends to zero for $v\rightarrow -\rho$. This corresponds to the presence of only one peak in FIG.~\ref{fig:1} which we emphasize by using a logarithmic scale on the $y$-axis \footnote{We use log-scale in all figures to emphasize the peaks and dips. The black points are given by the numerical simulation of the quantum walk. The full red curves always correspond to the approximation of the probability distribution with the limit density given by (\ref{dist:approx}).}.

Let us point out that the limit density of any homogeneous two-state quantum walk on a line can be transformed into the form (\ref{limit:dens:1/2}). Indeed, the two-dimensional Wigner coins ${\hat R}^{(1/2)}(\alpha,\beta,\gamma)$ covers all SU(2) matrices and the equivalence of a general SU(2) walk to the walk with the coin ${\hat R}^{(1/2)}(\rho)$ was already pointed out in \cite{unitary:equivalence}. Hence, for any two-state quantum walk on a line one can find two orthogonal coin states giving rise to non-generic distributions with only one peak.

Finally, let us rewrite the suitable basis $\left\{|\chi^\pm\rangle\right\}$ in terms of the eigenvectors of the coin operator (\ref{coin1/2}). This decomposition will be useful in the following Sections where we treat higher-dimensional models. The eigenvectors of the coin operator (\ref{coin1/2}) are given by
\begin{equation}
\label{eigenvectors1/2}
|\psi^\pm\rangle = \frac{1}{\sqrt{2}}(|1/2\rangle \mp i|-1/2\rangle).
\end{equation}
They satisfy the eigenvalue equations
$$
{\hat R}^{(1/2)}(\rho)|\psi^\pm\rangle = e^{\pm i\varphi}|\psi^\pm\rangle,
$$
where the phase of the eigenvalues reads
$$
\varphi = \arccos{\rho}.
$$
The decomposition of the suitable basis $\left\{|\chi^\pm\rangle\right\}$ into the eigenvectors (\ref{eigenvectors1/2}) is easily found to be
\begin{eqnarray}
\label{optimalEigen1/2}
\nonumber |\chi^+\rangle & = & \frac{1}{\sqrt{2}}\left(e^{-i\frac{\varphi}{2}}|\psi^+\rangle + e^{i\frac{\varphi}{2}}|\psi^-\rangle\right),\\
|\chi^-\rangle & = & \frac{i}{\sqrt{2}}\left(e^{-i\frac{\varphi}{2}}|\psi^+\rangle - e^{i\frac{\varphi}{2}}|\psi^-\rangle\right).
\end{eqnarray}
These relations provide us with a recipe for construction of suitable bases for $j>\frac{1}{2}$.

\section{Three-state model}
\label{sec:1}

In this Section we analyze the three-state Wigner walk which is driven by the coin operator
\begin{widetext}
$$
R^{(1)}(\rho) = \left(
            \begin{array}{ccc}
              \rho^2 & -\sqrt{2}\rho\sqrt{1-\rho^2} & 1-\rho^2 \\
              \sqrt{2}\rho\sqrt{1-\rho^2} & -1+2\rho^2 & -\sqrt{2}\rho\sqrt{1-\rho^2} \\
              1-\rho^2 & \sqrt{2}\rho\sqrt{1-\rho^2} & \rho^2 \\
            \end{array}
          \right).
$$
\end{widetext}
The Wigner matrix $R^{(1)}(\rho)$ is reminiscent of the modified Grover coin which was introduced in \cite{stef:cont:def} and recently analyzed in more detail in \cite{stef:limit,machida}. Therefore, it is not surprising that the two models yield the same results, as we identify in the following subsections. In fact, the affinity of the three-state Wigner walk and the three-state Grover walk was already discussed in \cite{konno:wigner}.

In the three-state Wigner walk the particle is allowed to remain at its present position. As we have discussed before, this implies that the probability distribution of the three-state Wigner walk is not described solely by the limit density (\ref{totaldens}). There is an additional non-vanishing and stationary peak at the origin due to the trapping effect (\ref{trap:prob}). We analyze the limit density in subsection \ref{sec:1:a} and the trapping probability in subsection \ref{sec:1:b}. Before we give their explicit forms, we first construct the suitable basis of the coin space. We adopt the recipe provided by Eq.~(\ref{optimalEigen1/2}), which gives the decomposition of the suitable basis states for the two-state model into the eigenvectors of the coin operator. For the three-state model, the coin ${\hat R}^{(1)}(\rho)$ has two eigenvectors $|\psi^\pm\rangle$ satisfying the eigenvalue equations
$$
{\hat R}^{(1)}(\rho)|\psi^\pm\rangle = e^{\pm i\varphi}|\psi^\pm\rangle.
$$
The explicit form of the eigenvectors and the phase $\varphi$ is left for the Appendix~\ref{app:3state}. Following the formula (\ref{optimalEigen1/2}) for the two-state model, we construct from $|\psi^\pm\rangle$ two orthonormal vectors $|\chi^\pm\rangle$ which will serve as part of the new basis. Moreover, the coin ${\hat R}^{(1)}(\rho)$ has an additional eigenvector $|\psi_0\rangle$ corresponding to the eigenvalue one. This vector is indeed orthogonal to both $|\psi^\pm\rangle$ and $|\chi^\pm\rangle$. Hence, we consider it as the last vector of the new basis which reads
\begin{eqnarray}
\label{basis:1}
\nonumber |\chi_0\rangle & = & |\psi_0\rangle, \\
\nonumber |\chi^+\rangle & = & \frac{1}{\sqrt{2}}\left(e^{-i\frac{\varphi}{2}}|\psi^+\rangle + e^{i\frac{\varphi}{2}}|\psi^-\rangle\right), \\
|\chi^-\rangle & = & \frac{i}{\sqrt{2}}\left(e^{-i\frac{\varphi}{2}}|\psi^+\rangle - e^{i\frac{\varphi}{2}}|\psi^-\rangle\right).
\end{eqnarray}
The initial state is now decomposed in the suitable basis $\{|\chi_0\rangle,|\chi^\pm\rangle\}$ according to
\begin{equation}
\label{InStOptimal1}
|\psi_C\rangle = h_0|\chi_0\rangle + h^+|\chi^+\rangle + h^-|\chi^-\rangle.
\end{equation}
The explicit correspondence between the amplitudes of the initial state in the suitable basis $h_i$ and in the standard basis $q_i$ is given in the Appendix~\ref{app:3state}. In the following we show that the limit density and the trapping probability obtain much more convenient forms when expressed in the basis $\{|\chi_0\rangle,|\chi^\pm\rangle \}$.

\subsection{Limit density}
\label{sec:1:a}

The limit density for the three-state Wigner walk was derived in \cite{konno:wigner} and reads
$$
\nu^{(1)}(v) = \frac{1}{2}\mu\left(\frac{v}{2};\rho\right){\cal M}^{(1,1)}\left(\frac{v}{2}\right),
$$
where ${\cal M}^{(1,1)}(v)$ can be expressed as a polynomial of degree two in $v$
$$
{\cal M}^{(1,1)}(v) = {\cal M}_0^{(1,1)} + {\cal M}_1^{(1,1)} v + {\cal M}_2^{(1,1)}v^2.
$$
The individual terms ${\cal M}_i^{(1,1)}$ depend on the coin parameter $\rho$ and the initial coin state. Their explicit form in the standard basis was given in \cite{konno:wigner}, we present it in the Appendix~\ref{app:3state} for comparison. There we also show how the terms ${\cal M}_i^{(1,1)}$ simplify when we expand the initial state into the suitable basis according to (\ref{InStOptimal1}). Indeed, in the suitable basis we obtain
\begin{eqnarray}
\nonumber {\cal M}_0^{(1,1)} & = & |h^+|^2 + |h^-|^2,\\
\nonumber {\cal M}_1^{(1,1)} & = & \frac{1}{\rho}(h_0\overline{h^-} + \overline{h_0} h^-),\\
\nonumber {\cal M}_2^{(1,1)} & = & \frac{1}{\rho^2}(|h_0|^2 - |h^+|^2).
\end{eqnarray}

\begin{figure}[htbp]
\begin{center}
\includegraphics[width=0.45\textwidth]{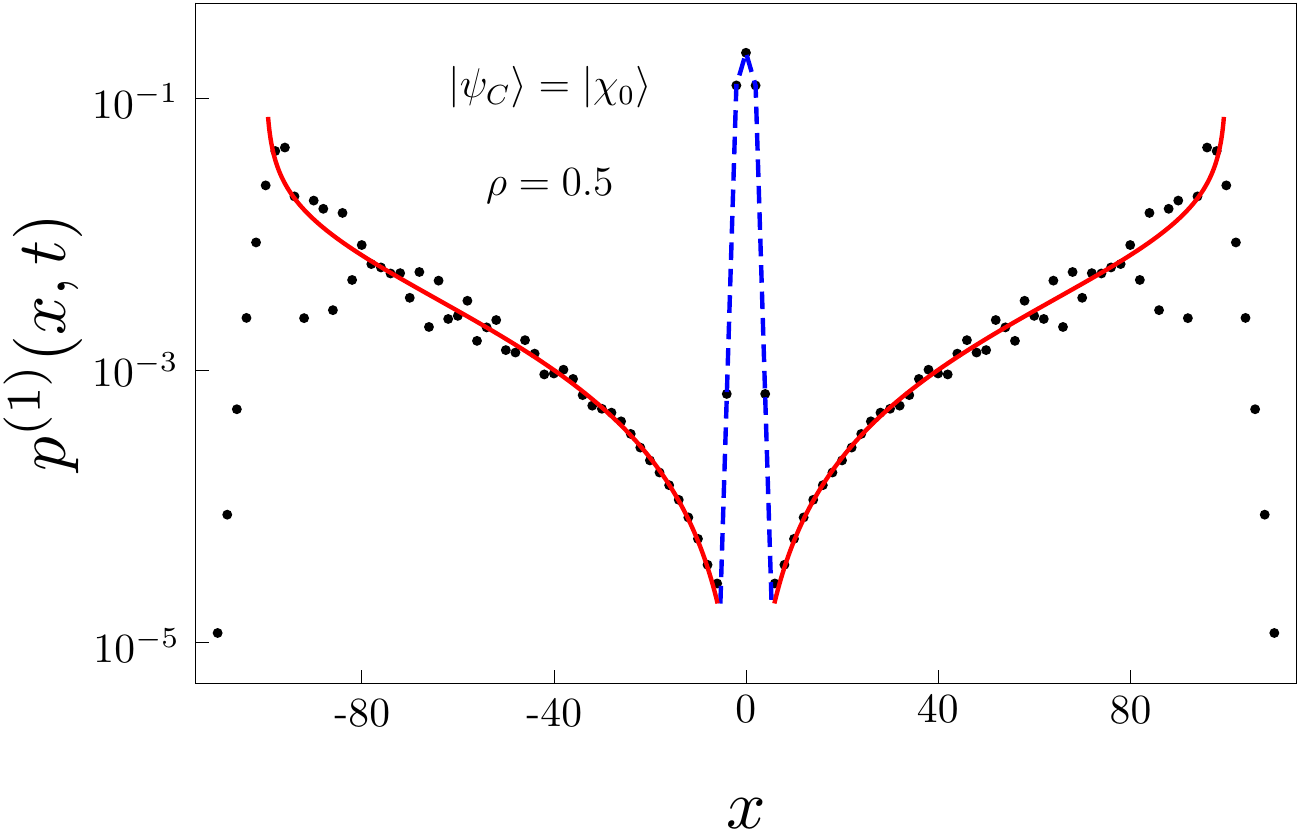}
\end{center}
\caption{Probability distribution after 100 steps of the three-state Wigner walk. The initial coin state was chosen as $|\chi_0\rangle$ and the coin parameter is $\rho=0.5$ }
\label{fig:1:h0}
\end{figure}

\begin{figure}[htbp]
\begin{center}
\includegraphics[width=0.45\textwidth]{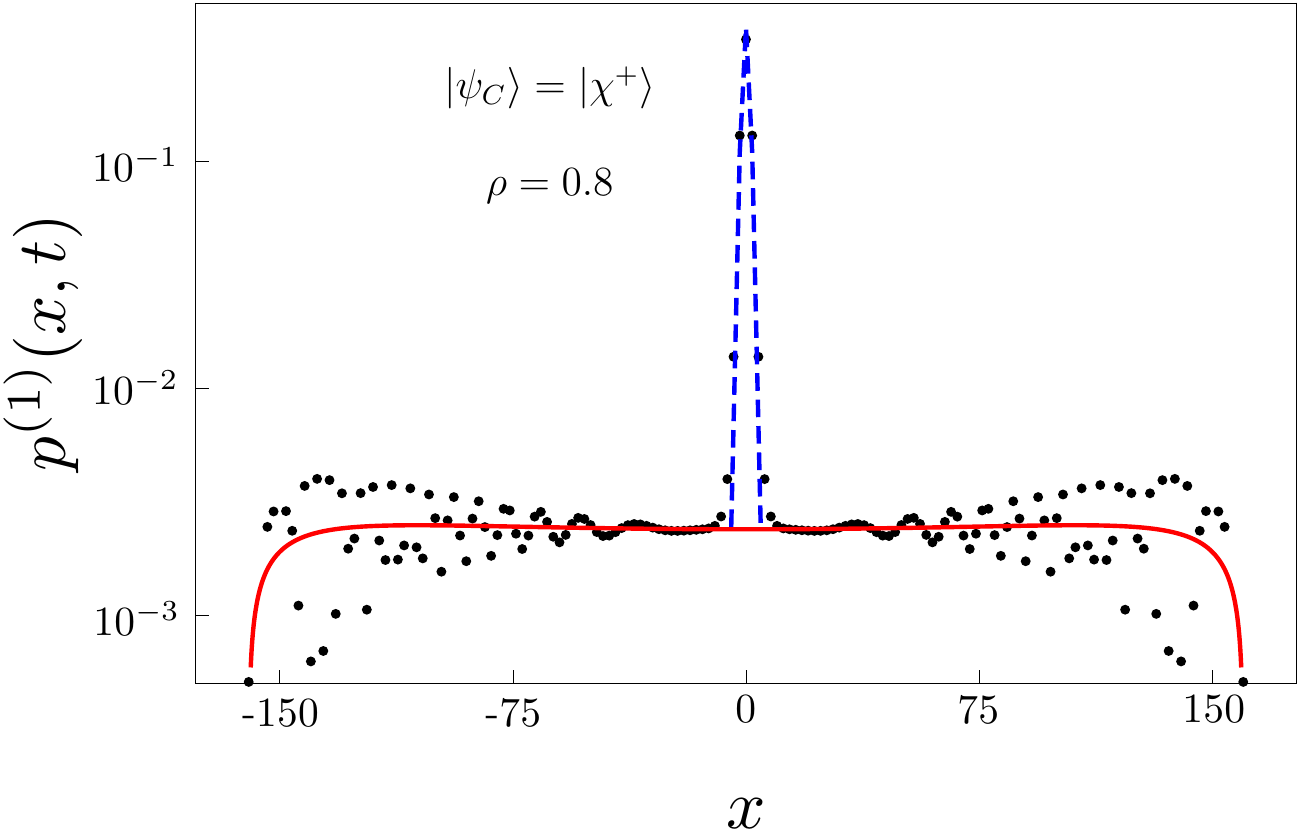}
\end{center}
\caption{Probability distribution after 100 steps of the three-state Wigner walk. The initial coin state was chosen as $|\chi^+\rangle$ and the coin parameter is $\rho=0.8$ }
\label{fig:1:hp}
\end{figure}

\begin{figure}[htbp]
\begin{center}
\includegraphics[width=0.45\textwidth]{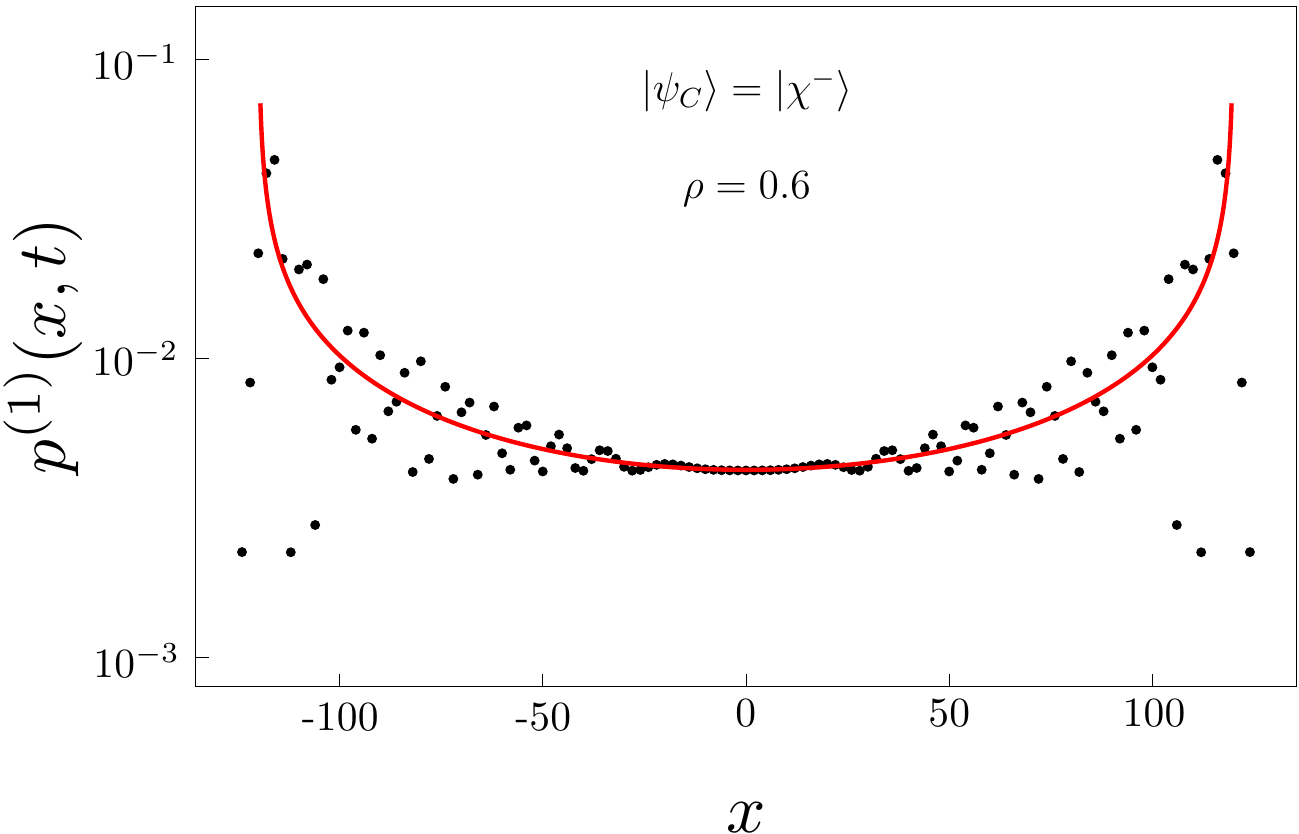}
\end{center}
\caption{Probability distribution after 100 steps of the three-state Wigner walk. The initial coin state was chosen as $|\chi^-\rangle$ and the coin parameter is $\rho=0.6$ }
\label{fig:1:hm}
\end{figure}

For illustration we show in FIGs.~\ref{fig:1:h0},\ref{fig:1:hp} and \ref{fig:1:hm} the probability distribution of the three-state Wigner walk with the initial coin state given by one of the suitable basis vectors (\ref{basis:1}). In FIG.~\ref{fig:1:h0} we find that for $|\chi_0\rangle$ the limit density tends to zero at the origin. Indeed, for $h_0=1$ the constant term of the limit density ${\cal M}_0^{(1,1)}$ vanishes. Nevertheless, the probability distribution at the origin does not vanish due to the trapping effect\footnote{We calculate the explicit form of the trapping probability in the following subsection. The trapping probability is in this and all relevant figures denoted by the dashed blue curve.}. For $|\chi^+\rangle$ both peaks at $v\rightarrow\pm 2\rho$ vanishes, as we illustrate in FIG.~\ref{fig:1:hp}. Finally, FIG.~\ref{fig:1:hm} indicates that for the last basis vector $|\chi^-\rangle$ the trapping at the origin disappears, which we identify analytically in the Appendix~\ref{app:3state}.

We point out that the limit density of the three-state Grover walk and its one-parameter extension exhibit all of these features \cite{stef:limit}. Moreover, in the following subsection we show that also the trapping effect of the three-state Wigner walk is the same as for the three-state Grover walk.

\subsection{Trapping probability}
\label{sec:1:b}

Let us now turn to the trapping probability. The details of the calculations are left for the Appendix~\ref{app:3state}. Following the same approach that was used for the three-state Grover walk in \cite{falkner,machida,stef:limit} we find that in the suitable basis (\ref{basis:1}) the trapping probability reads
\begin{equation}
\label{trapping1}
p^{(1)}_\infty(2x) = \left\{
                \begin{array}{c}
                  Q^{2|x|}\frac{2(1-\rho^2)}{\rho^4}|h_0 - h^+|^2,\quad x < 0, \\
                  \ \\
                  \frac{Q}{\rho^2}((1-\rho^2)|h_0|^2 + |h^+|^2),\quad x = 0, \\
                  \ \\
                  Q^{2x}\frac{2(1-\rho^2)}{\rho^4}|h_0 + h^+|^2,\quad x > 0.
                \end{array}
              \right.
\end{equation}
Here we have denoted
\begin{equation}
\label{Q}
Q = \frac{2-\rho^2 - 2\sqrt{1-\rho^2}}{\rho^2}.
\end{equation}
We note that the trapping probability for the one-parameter extension of the three-state Grover walk has exactly the same form \cite{stef:limit}.

\begin{figure}[htbp]
\begin{center}
\includegraphics[width=0.45\textwidth]{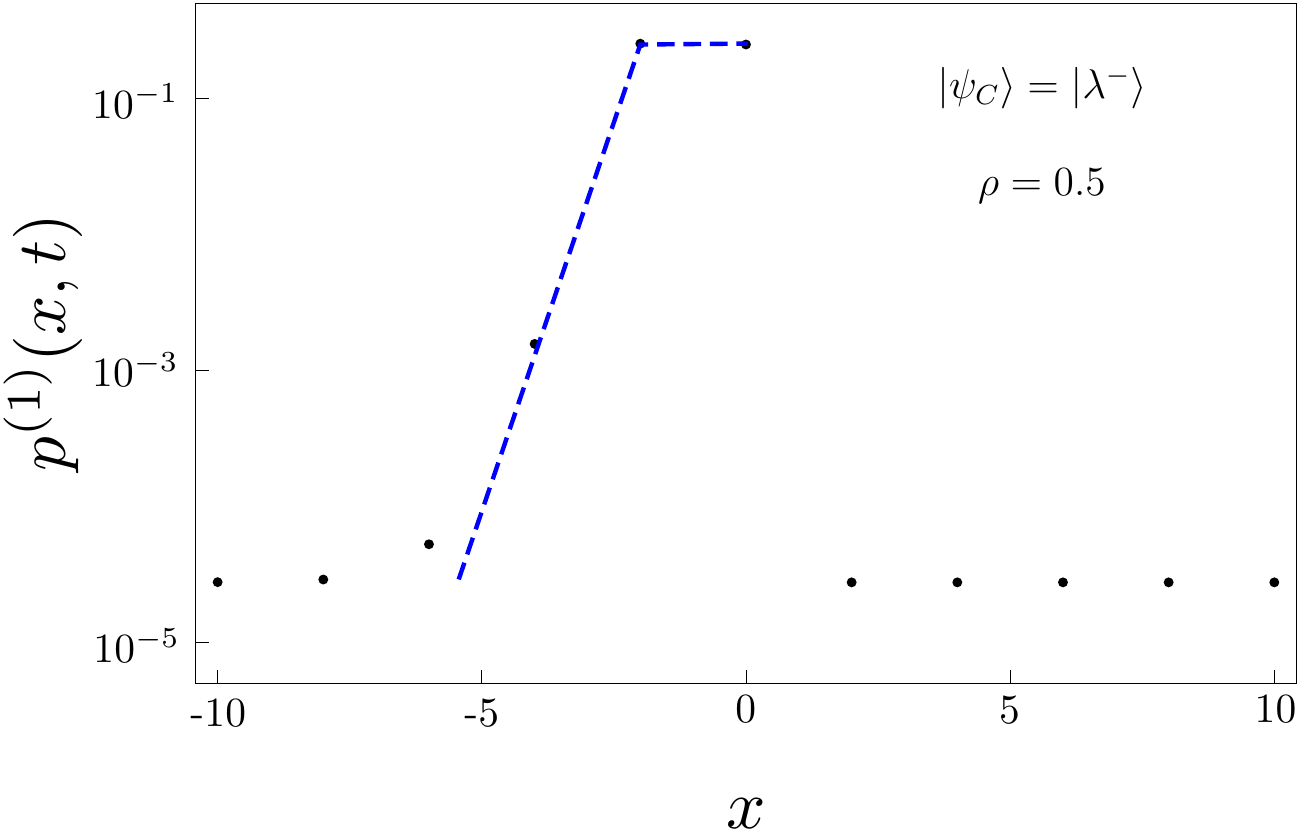}
\end{center}
\caption{Probability distribution after 10000 steps of the three-state Wigner walk. The initial coin state was chosen as $|\lambda^-\rangle$ and the coin parameter is $\rho=0.5$. Only a small vicinity of the origin is displayed to emphasize that for $|\lambda^-\rangle$ the trapping effect appears only for $x\leq 0$. Higher number of steps in comparison with other figures was chosen so that the trapping effect is sufficiently pronounced further from the origin.}
\label{fig:1:lm}
\end{figure}

In the suitable basis (\ref{basis:1}), the trapping probability (\ref{trapping1}) is independent of the amplitude $h^-$ of the initial coin state (\ref{InStOptimal1}). This fact reduces the dependence of the trapping probability to just two amplitudes $h_0$ and $h^+$. However, the expression (\ref{trapping1}) can be simplified further by an additional change of basis. Notice that the dependence of the trapping probability is different for positive and negative positions. In fact, one can choose such an initial state that the trapping effect appears only on positive or negative semi-axis. We note that the same feature was identified for the three-state Grover walk \cite{falkner} and its one-parameter extension \cite{stef:limit}. For the three-state Wigner walk with the initial coin state
$$
|\lambda^+\rangle = \frac{1}{\sqrt{2}}\left(|\chi_0\rangle + |\chi^+\rangle\right),
$$
the trapping effect appears only for non-negative positions. Similarly, the state
$$
|\lambda^-\rangle = \frac{1}{\sqrt{2}}\left(|\chi_0\rangle - |\chi^+\rangle\right)
$$
shows trapping only for non-positive $x$. We illustrate this feature in FIG.~\ref{fig:1:lm}.

The vectors $|\lambda^+\rangle$ and $|\lambda^-\rangle$ are mutually orthogonal. Moreover, they are both orthogonal to $|\chi^-\rangle$ and the triplet $\{|\chi^-\rangle,|\lambda^\pm\rangle \}$ forms an orthonormal basis of the coin space. When we decompose the initial coin state as
$$
|\psi_C\rangle = h^-|\chi^-\rangle + l^+|\lambda^+\rangle + l^-|\lambda^-\rangle,
$$
we find that the trapping probability turns into
$$
p_\infty(2x) = \left\{
                \begin{array}{c}
                  Q^{2|x|}\frac{2(1-\rho^2)}{\rho^4}|l^-|^2,\quad x < 0, \\
                  \ \\
                  \frac{Q}{\rho^2}\left(|l^+|^2 + |l^-|^2 - \frac{1}{2\rho^2}|l^+ + l^-|^2\right),\quad x = 0, \\
                  \ \\
                  Q^{2x}\frac{2(1-\rho^2)}{\rho^4}|l^+|^2,\quad x > 0.
                \end{array}
              \right.
$$
The advantage of the basis $\{|\chi^-\rangle,|\lambda^\pm\rangle \}$ is that the trapping probability outside the origin has a simpler form. Namely, it depends only on one amplitude $l^+$ (resp. $l^-$) for positive $x$ (resp. negative $x$). This additional change of basis becomes crucial in Section~\ref{sec:2} where we treat the five-state Wigner walk.

\section{Four-state model}
\label{sec:3/2}

For $j=\frac{3}{2}$ we obtain a four state quantum walk model with the coin operator determined by (\ref{matrix:rho}). The coin operator ${\hat R}^{(3/2)}(\rho)$ has two pairs of eigenvectors $|\psi_1^\pm\rangle$ and $|\psi_2^\pm\rangle$ with conjugated pairs of eigenvalues
\begin{eqnarray}
\nonumber {\hat R}^{(3/2)}(\rho)|\psi_1^\pm\rangle & = & e^{\pm i\varphi_1}|\psi_1^\pm\rangle,\\
\nonumber {\hat R}^{(3/2)}(\rho)|\psi_2^\pm\rangle & = & e^{\pm i\varphi_2}|\psi_2^\pm\rangle.
\end{eqnarray}
The explicit form of the eigenvectors and the phases $\varphi_{1,2}$ is given in the Appendix~\ref{app:4state}. Following the formula (\ref{optimalEigen1/2}) we construct the suitable basis by combining only the eigenvectors corresponding to the same phase factor $\varphi_1$ or $\varphi_2$. Therefore, we consider the suitable basis in the form
\begin{eqnarray}
\label{basis:3:2}
\nonumber |\chi_1^+\rangle & = & \frac{1}{\sqrt{2}} (e^{-i\frac{\varphi_1}{2}}|\psi_1^+\rangle + e^{i\frac{\varphi_1}{2}}|\psi_1^-\rangle), \\
\nonumber  |\chi_1^-\rangle & = & \frac{i}{\sqrt{2}} (e^{-i\frac{\varphi_1}{2}}|\psi_1^+\rangle - e^{i\frac{\varphi_1}{2}}|\psi_1^-\rangle ), \\
\nonumber |\chi_2^+\rangle & = & \frac{1}{\sqrt{2}} (e^{-i\frac{\varphi_2}{2}}|\psi_2^+\rangle + e^{i\frac{\varphi_2}{2}}|\psi_2^-\rangle),\\
|\chi_2^-\rangle & = & \frac{i}{\sqrt{2}} (e^{-i\frac{\varphi_2}{2}}|\psi_2^+\rangle - e^{i\frac{\varphi_2}{2}}|\psi_2^-\rangle).
\end{eqnarray}
The initial coin state is decomposed in the suitable basis according to
$$
|\psi_C\rangle = h_1^+|\chi_1^+\rangle + h_1^-|\chi_1^-\rangle + h_2^+|\chi_2^+\rangle + h_2^-|\chi_2^-\rangle .
$$
We present the explicit relation between the amplitudes in the suitable basis $h_i$ and the standard basis $q_i$ in the Appendix~\ref{app:4state}.

\begin{figure}[htbp]
\begin{center}
\includegraphics[width=0.45\textwidth]{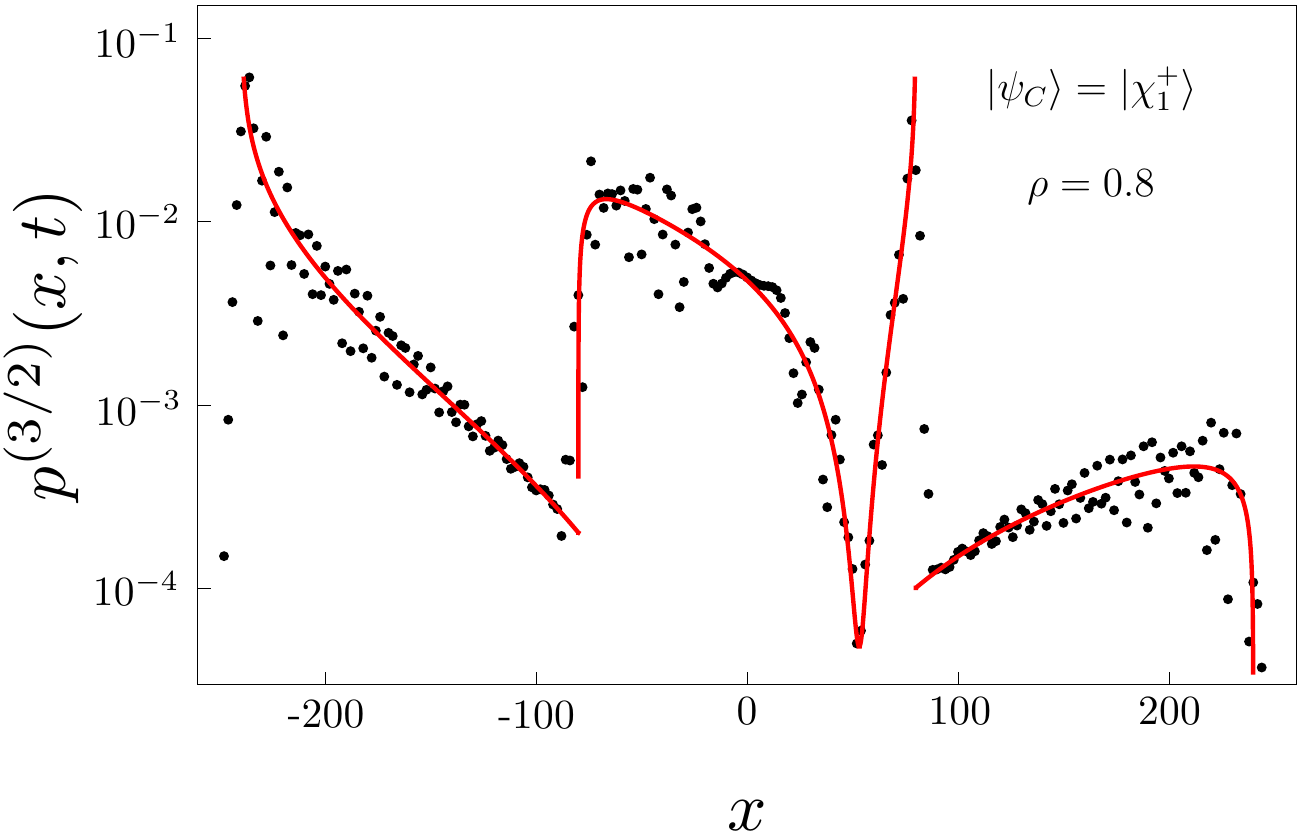}
\end{center}
\caption{(Color
  online) Probability distribution after 100 steps of the four-state Wigner walk. The initial coin state was chosen as $|\chi_1^+\rangle$. The coin parameter is $\rho=0.8$. The initial state $|\chi_1^-\rangle$ results in a distribution which is a mirror image of the present one.}
\label{fig:32:h1}
\end{figure}

\begin{figure}[htbp]
\begin{center}
\includegraphics[width=0.45\textwidth]{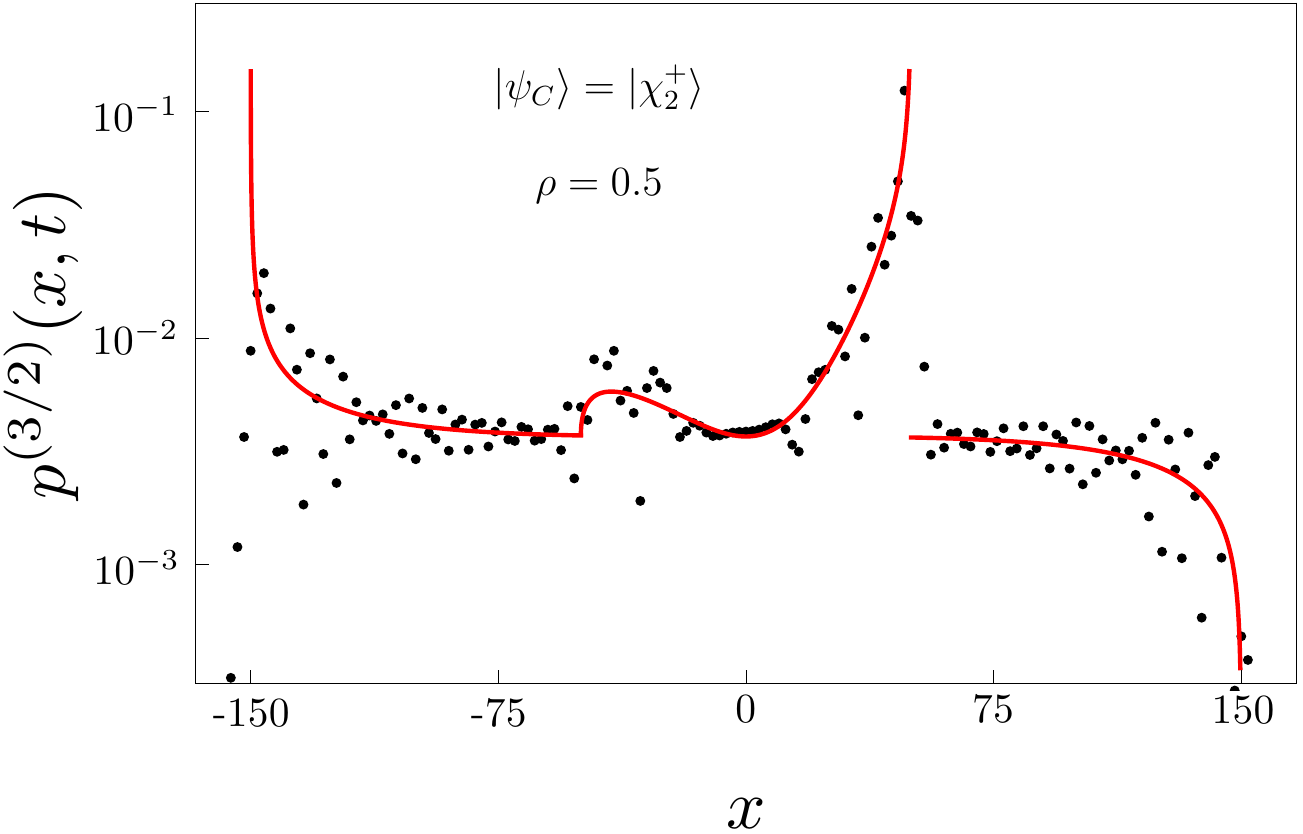}
\end{center}
\caption{(Color
  online) Probability distribution after 100 steps of the four-state Wigner walk. The initial coin state was chosen as $|\chi_2^+\rangle$. The coin parameter is $\rho=0.5$. For the initial state $|\chi_2^-\rangle$ the probability distribution is a mirror image of the present one.}
\label{fig:32:h2}
\end{figure}

Let us now turn to the limit density which for the four-state model is given by a sum of two densities \cite{konno:wigner}
$$
\nu^{(3/2)}(v) = \nu^{(3/2,1/2)}(v) + \nu^{(3/2,3/2)}(v).
$$
The individual densities corresponding to a slower walk ($m=\frac{1}{2}$) and a faster walk ($m=\frac{3}{2}$) read
\begin{eqnarray}
\nonumber \nu^{(3/2,1/2)}(v) & = & \mu(v;\rho){\cal M}^{(3/2,1/2)}(v), \\
\nu^{(3/2,3/2)}(v) & = & \frac{1}{3}\mu\left(\frac{v}{3};\rho\right){\cal M}^{(3/2,3/2)}\left(\frac{v}{3}\right),
\label{density3/2m1/2a3/2}
\end{eqnarray}
The weight functions ${\cal M}^{(3/2,m)}(v)$ are given by cubic polynomials in $v$
\begin{eqnarray}
\label{weight:3/2}
{\cal M}^{(3/2,m)}(v) & = & \sum_{k=0}^3{\cal M}_i^{(3/2,m)} v^i ,
\end{eqnarray}
with coefficients ${\cal M}_i^{(3/2,m)}$ determined by the initial coin state and the coin parameter $\rho$. Their explicit forms in the standard basis of the coin space are given in \cite{konno:wigner}. We express them in terms of the suitable basis $\{|\chi_1^\pm\rangle, |\chi_2^\pm\rangle\}$ in the Appendix~\ref{app:4state}. Using the expressions (\ref{weight:3/2:1/2}) and (\ref{weight:3/2:3/2}) one can show by direct computation that each of the basis states $|\chi_i^\pm\rangle$ eliminates two peaks of the probability distribution - one in each of the individual limit densities $\nu^{(3/2,1/2)}(v)$ and $\nu^{(3/2,3/2)}(v)$. To illustrate our results we present in FIGs.~\ref{fig:32:h1} and \ref{fig:32:h2} the probability distributions for the initial states $|\chi_1^+\rangle$ and $|\chi_2^+\rangle$.

\begin{figure}[htbp]
\begin{center}
\includegraphics[width=0.45\textwidth]{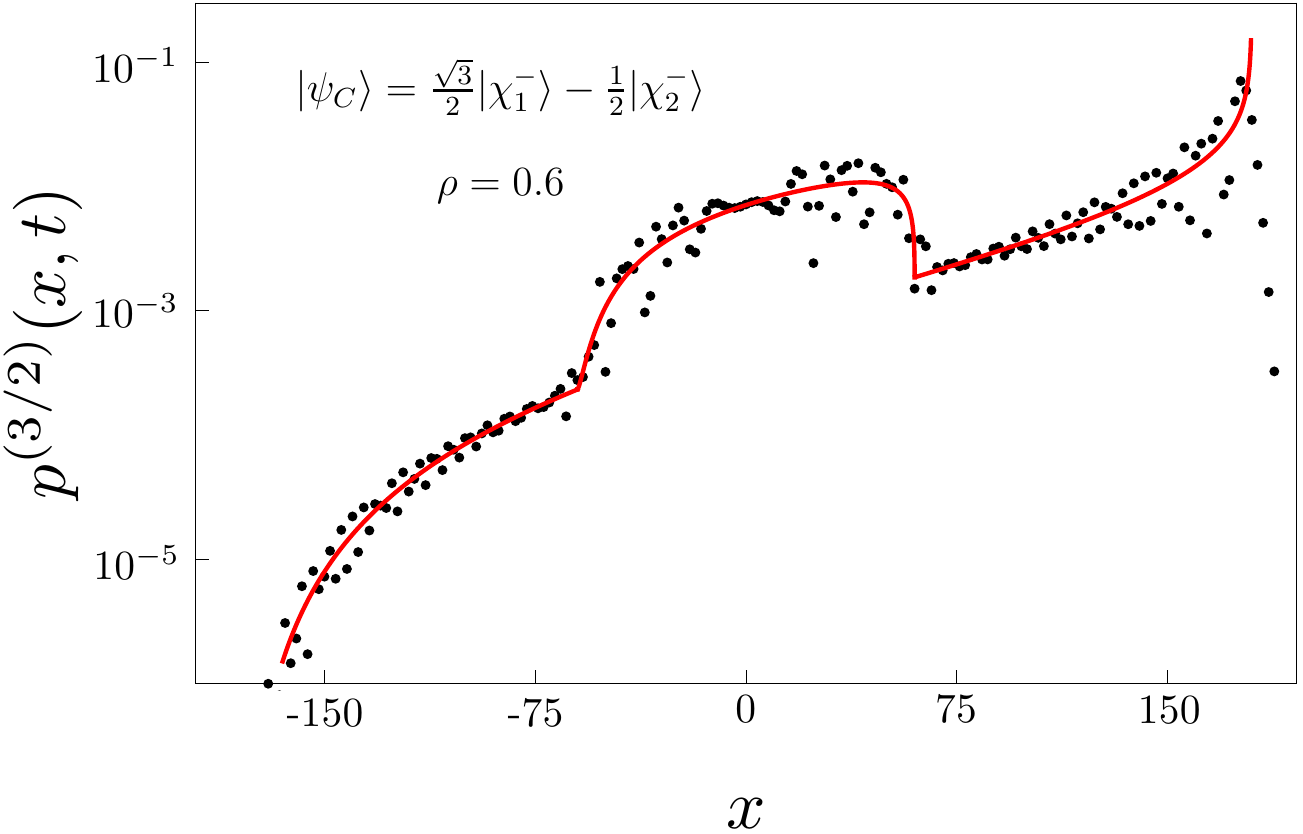}
\end{center}
\caption{(Color
  online) Probability distribution after 100 steps of the four-state Wigner walk. The initial coin state $\frac{\sqrt{3}}{2}|\chi_1^-\rangle - \frac{1}{2}|\chi_2^-\rangle$ was chosen as a part of the set (\ref{InStInnerPeaks}) for which the inner peaks in the probability distribution vanishes. Moreover, since $h_1^+=h_2^+=0$, only one of the outer peaks is present. The coin parameter is $\rho=0.6$ }
\label{fig:inner:peak}
\end{figure}

We point out that it is possible to eliminate both peaks for each limit density $\nu^{(3/2,m)}(v)$ on its own. In order to do so we have to find an initial coin state such that both divergences provided by the Konno's density $\mu(v;\rho)$ in (\ref{density3/2m1/2a3/2}) vanishes. Hence, the weight function have to be of the form
$$
{\cal M}^{(3/2,m)}(v) = (\rho^2-v^2)(a + b v),
$$
for some arbitrary constants $a$ and $b$. This is satisfied provided that
\begin{eqnarray}
\label{condElimj3/2}
\nonumber {\cal M}_0^{(3/2,m)} + \rho^2{\cal M}_2^{(3/2,m)} & = & 0,\\
 {\cal M}_1^{(3/2,m)} + \rho^2{\cal M}_3^{(3/2,m)} & = & 0.
\end{eqnarray}
Adding and subtracting these two equations for the weight function ${\cal M}^{(3/2,1/2)}(v)$ we obtain the following
$$
|h_1^++\sqrt{3} h_2^+|^2 = 0,\quad |h_1^-+\sqrt{3} h_2^-|^2 = 0,
$$
which lead us to the relations
\begin{equation}
\label{cond:slower}
h_2^+ = -\frac{1}{\sqrt{3}} h_1^+,\quad h_2^- = -\frac{1}{\sqrt{3}} h_1^-.
\end{equation}
Hence, we find a two-dimensional subset of states
\begin{equation}
\label{InStInnerPeaks}
|\psi_C^{(1/2)}\rangle = h_1^+|\chi_1^+\rangle + h_1^-|\chi_1^-\rangle - \frac{h_1^+}{\sqrt{3}}|\chi_2^+\rangle - \frac{h_1^-}{\sqrt{3}}|\chi_2^-\rangle,
\end{equation}
for which the peaks of the slower walk described by $\nu^{(3/2,1/2)}(v)$ vanishes. We note that for the vectors of the family (\ref{InStInnerPeaks}) with either $h_1^+=h_2^+=0$ or $h_1^-=h_2^-=0$ in addition one of the peaks of the faster walk vanishes, i.e. these states result in a probability distribution with only one peak. We illustrate this result in FIG.~\ref{fig:inner:peak}.

\begin{figure}[htbp]
\begin{center}
\includegraphics[width=0.45\textwidth]{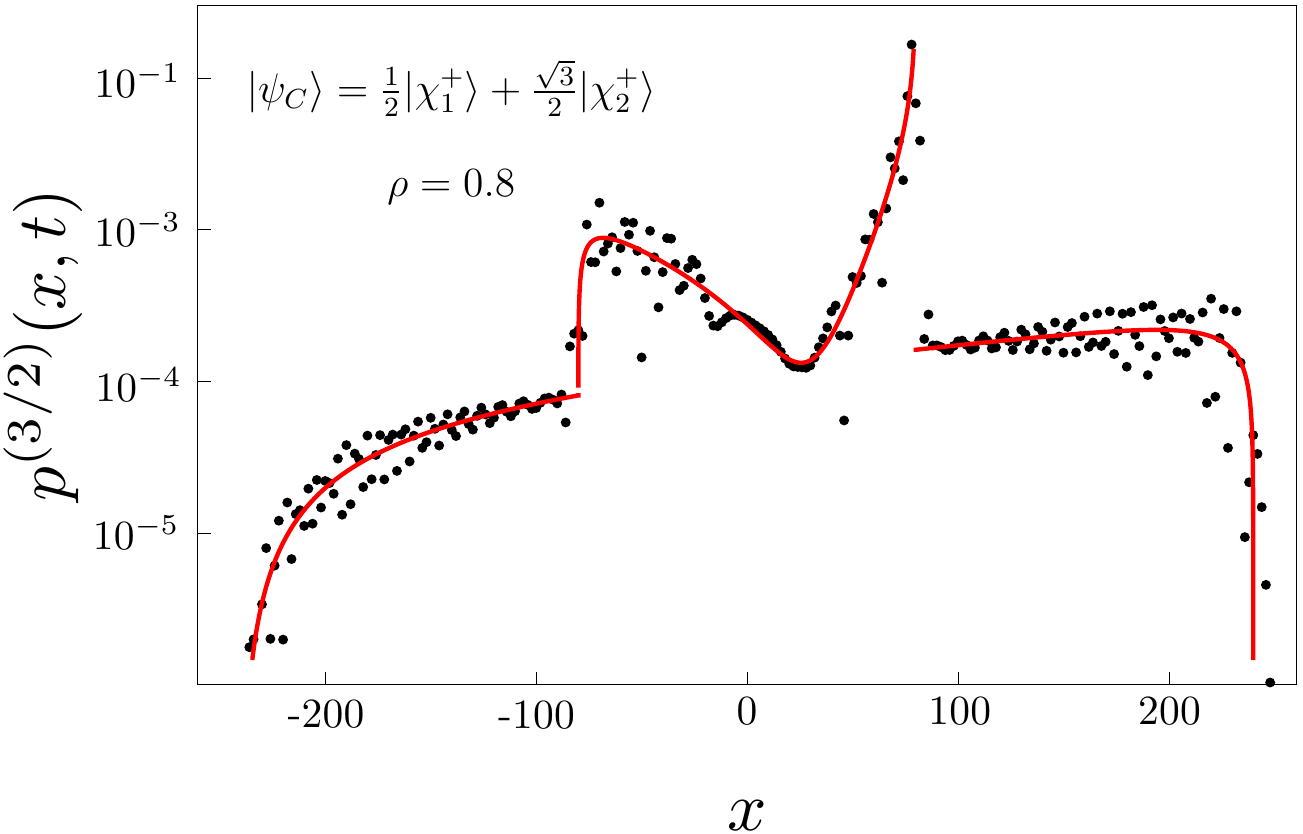}
\end{center}
\caption{(Color
  online) Probability distribution after 100 steps of the four-state Wigner walk. The initial coin state $\frac{1}{2}|\chi_1^+\rangle + \frac{\sqrt{3}}{2}|\chi_2^+\rangle$ is an element of the set (\ref{InStOuterPeaks}) for which the outer peaks in the probability distribution vanishes. Since $h_1^-=h_2^-=0$, only one of the inner peaks is present. The coin parameter is $\rho=0.8$ }
\label{fig:outer:peak}
\end{figure}

Similarly, we can cancel both divergencies for the second limit density $\nu^{(3/2,3/2)}(v)$ corresponding to the faster peaks in the probability distribution. From the equations (\ref{condElimj3/2}) for $m=\frac{3}{2}$ we find that both faster peaks vanish provided that
\begin{equation}
\label{condOutPeaks}
h_2^+ = \sqrt{3} h_1^+,\quad h_2^- = \sqrt{3} h_1^-.
\end{equation}
Hence, we obtain a two-dimensional set of initial states
\begin{equation}
\label{InStOuterPeaks}
|\psi_C^{(3/2)}\rangle = h_1^+|\chi_1^+\rangle + h_1^-|\chi_1^-\rangle + \sqrt{3} h_1^+|\chi_2^+\rangle + \sqrt{3}h_1^-|\chi_2^-\rangle,
\end{equation}
for which the outer peaks in the probability distribution vanishes. Moreover, the vectors from the family (\ref{InStOuterPeaks}) satisfying either $h_1^+=h_2^+=0$ or $h_1^-=h_2^-=0$ lead to elimination of one additional peak of the slower walk, i.e. they result in a probability distribution with only one peak. We illustrate this result in FIG.~\ref{fig:outer:peak}.

Clearly, it is impossible to fulfill both conditions (\ref{cond:slower}) and (\ref{condOutPeaks}) simultaneously. Hence, the probability distribution of the four-state Wigner walk has always at least one peak.

\section{Five-state model}
\label{sec:2}

In this Section we analyze the case $j=2$ which leads to the five-state Wigner walk. Similarly to the three-state model, which we have discussed in Section~\ref{sec:1}, the probability distribution of the five-state model consists of the limit density and the trapping probability. We treat them in subsections \ref{sec:2a} and \ref{sec:2b}, but first we construct the suitable basis of the coin space following the recipe (\ref{optimalEigen1/2}). The coin operator ${\hat R}^{(2)}(\rho)$ has two pairs of eigenvectors $|\psi_1^\pm\rangle$ and $|\psi_2^\pm\rangle$ corresponding to two pairs of conjugated eigenvalues
\begin{eqnarray}
\nonumber {\hat R}^{(2)}(\rho)|\psi_1^\pm\rangle & = & e^{\pm i \varphi_1}|\psi_1^\pm\rangle, \\
\nonumber {\hat R}^{(2)}(\rho)|\psi_2^\pm\rangle & = & e^{\pm i \varphi_2}|\psi_2^\pm\rangle.
\end{eqnarray}
The explicit form of the eigenvectors and the phases $\varphi_{1,2}$ is left for the Appendix~\ref{app:5state}. From these two pairs of eigenvectors we construct four basis vectors $|\chi_1^\pm\rangle$ and $|\chi_2^\pm\rangle$ according to (\ref{basis:3:2}). In addition, the coin ${\hat R}^{(2)}(\rho)$ has an eigenvalue 1.  We include the corresponding eigenvector $|\psi_0\rangle$ as the last basis vector. Hence, we obtain the suitable basis in the form
\begin{eqnarray}
\label{basis:2}
\nonumber |\chi_0\rangle & = & |\psi_0\rangle,\\
\nonumber |\chi_1^+ \rangle & = & \frac{1}{\sqrt{2}}\left(e^{-i\frac{\varphi_1}{2}}|\psi_1^+\rangle + e^{i\frac{\varphi_1}{2}}|\psi_1^-\rangle\right), \\
\nonumber |\chi_1^- \rangle & = & \frac{i}{\sqrt{2}}\left(e^{-i\frac{\varphi_1}{2}}|\psi_1^+\rangle - e^{i\frac{\varphi_1}{2}}|\psi_1^-\rangle\right), \\
\nonumber |\chi_2^+ \rangle & = & \frac{1}{\sqrt{2}}\left(e^{-i\frac{\varphi_1}{2}}|\psi_2^+\rangle + e^{i\frac{\varphi_1}{2}}|\psi_2^-\rangle\right), \\
|\chi_2^- \rangle & = & \frac{i}{\sqrt{2}}\left(e^{-i\frac{\varphi_1}{2}}|\psi_2^+\rangle - e^{i\frac{\varphi_1}{2}}|\psi_2^-\rangle\right).
\end{eqnarray}
The initial state is decomposed into the suitable basis according to
$$
|\psi_C\rangle = h_0|\chi_0\rangle + h_1^+|\chi_1^+\rangle + h_1^-|\chi_1^-\rangle + h_2^+|\chi_2^+\rangle + h_2^-|\chi_2^-\rangle.
$$
We present the correspondence between the amplitudes in the suitable basis $h_i$ and in the standard basis $q_i$ in the Appendix~\ref{app:5state}. In the following subsections we illustrate that the change of the basis allows us to identify various interesting regimes of dynamics which are otherwise hidden in the standard basis description.

\subsection{Limit density}
\label{sec:2a}

Let us first discuss the limit density. As for the four-state walk of Section~\ref{sec:3/2}, the total limit density (\ref{totaldens}) is a sum of two densities
$$
\nu^{(2)}(v) = \nu^{(2,1)}(v) + \nu^{(2,2)}(v),
$$
corresponding to a slower walk
$$
\nu^{(2,1)}(v) = \frac{1}{2}\mu\left(\frac{v}{2};\rho\right){\cal M}^{(2,1)}\left(\frac{v}{2}\right)
$$
and a faster walk
$$
\nu^{(2,2)}(v) = \frac{1}{4}\mu\left(\frac{v}{4};\rho\right){\cal M}^{(2,2)}\left(\frac{v}{4}\right).
$$
The weight functions ${\cal M}^{(2,m)}(v)$ are polynomials of degree four in $v$
$$
{\cal M}^{(2,m)}(v) = \sum\limits_{i=0}^4 {\cal M}_i^{(2,m)}v^i,
$$
with coefficients ${\cal M}_i^{(2,m)}$ determined by the initial coin state and the coin parameter $\rho$. Their explicit form in the standard basis of the coin space can be evaluated using the procedure of \cite{konno:wigner}. In the Appendix~\ref{app:5state} we express them in terms of the coefficients in the suitable basis.

\begin{figure}[h!]
\begin{center}
\includegraphics[width=0.45\textwidth]{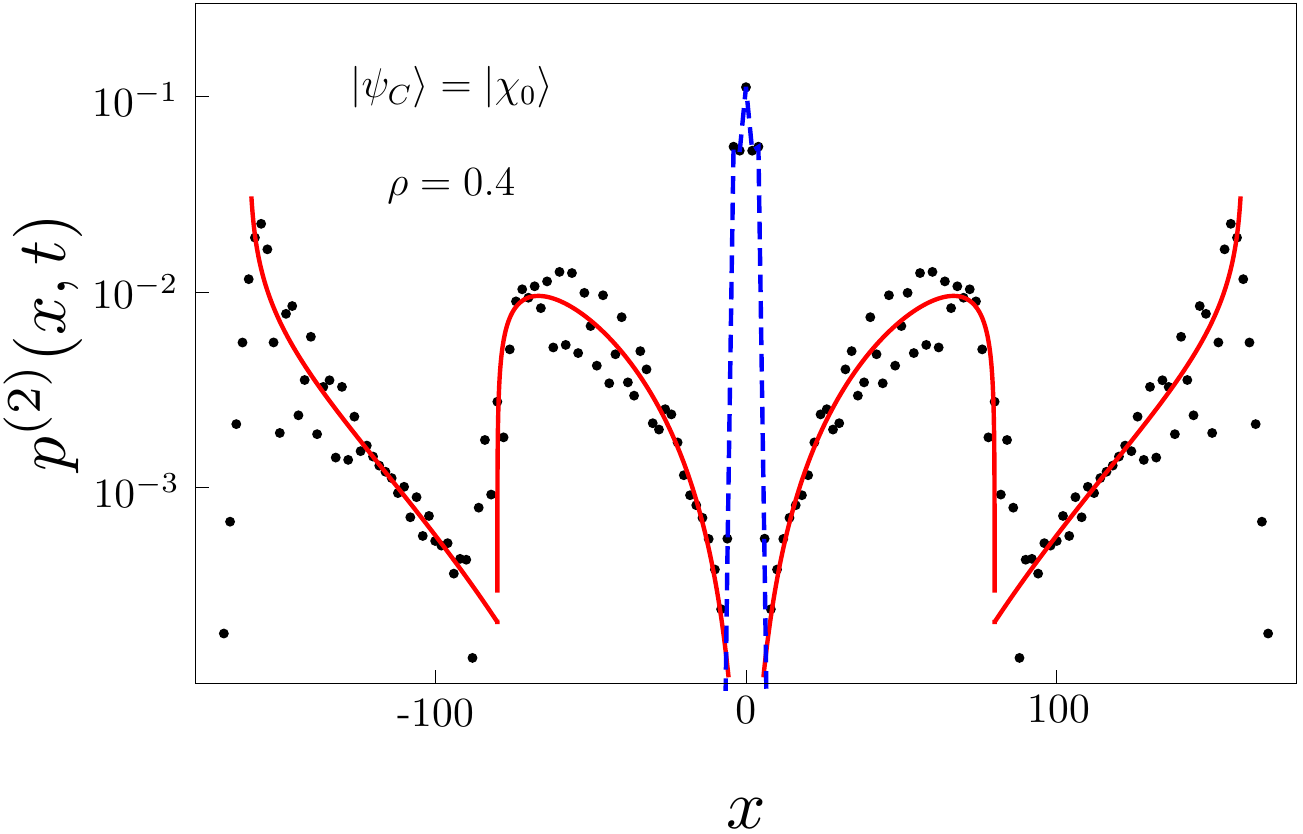}
\end{center}
\caption{(Color
  online) Probability distribution after 100 steps of the five-state Wigner walk. The initial coin state was chosen as $|\chi_0\rangle$. The coin parameter is $\rho=0.4$ }
\label{fig:2:h0}
\end{figure}

Let us now illustrate the role of individual vectors of the suitable basis $\{|\chi_0\rangle,|\chi_1^\pm\rangle,|\chi_2^\pm\rangle\}$ on the dynamics of the five-state Wigner walk. Using the explicit form of the weight functions (\ref{weight:2:1}) and (\ref{weight:2:2}) one can show that each of the basis states eliminates two peaks, either in the slower walk described $\nu^{(2,1)}(v)$ or in the faster walk given by $\nu^{(2,2)}(v)$. In FIG.~\ref{fig:2:h0} we display the probability distribution for the initial state $|\chi_0\rangle$. We find that both peaks of the slower walk vanishes. In addition, both densities $\nu^{(2,1)}(v)$ and $\nu^{(2,2)}(v)$ tend to zero at the origin, since both weight functions ${\cal M}^{(2,m)}(v)$ miss the constant term ${\cal M}^{(2,m)}_0$. In FIG.~\ref{fig:2:h0} this corresponds to the significant dip of the red curve around the origin. However, the total probability distribution does not vanish at the origin due to the trapping effect, which is illustrated by the blue curve in FIG.~\ref{fig:2:h0}.

\begin{figure}[h!]
\begin{center}
\includegraphics[width=0.45\textwidth]{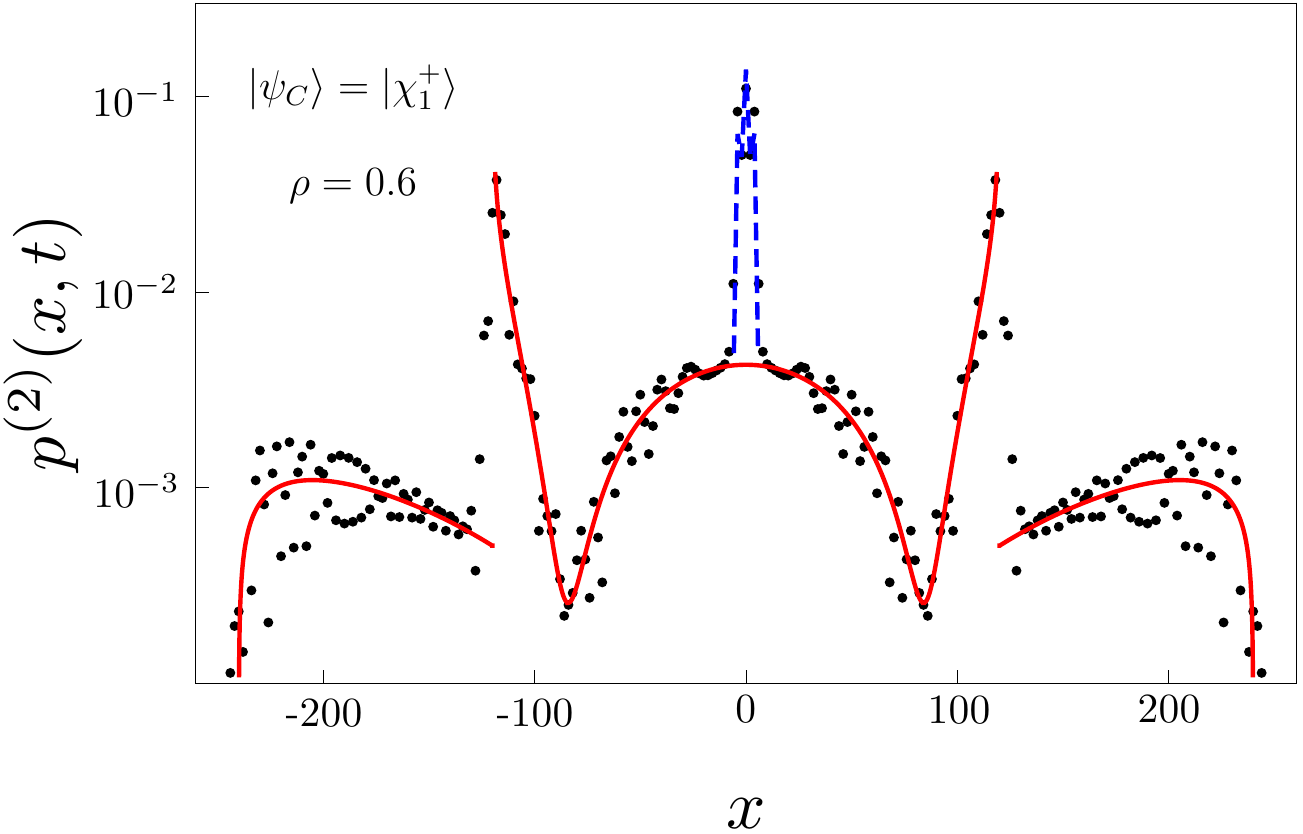}
\end{center}
\caption{(Color
  online) Probability distribution after 100 steps of the five-state Wigner walk. The initial coin state was chosen as $|\chi_1^+\rangle$. The coin parameter is $\rho=0.6$ }
\label{fig:2:h1p}
\end{figure}

\begin{figure}[h!]
\begin{center}
\includegraphics[width=0.45\textwidth]{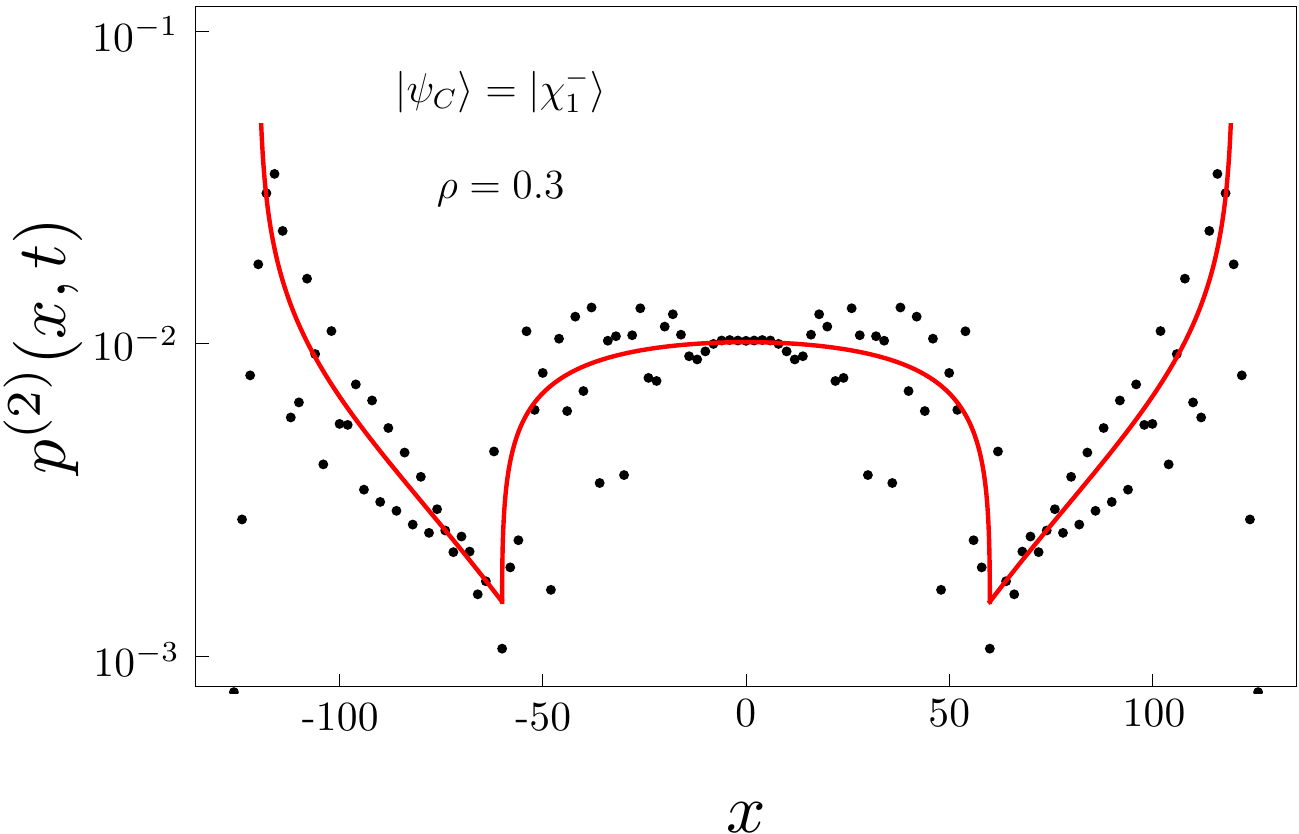}
\end{center}
\caption{(Color
  online) Probability distribution after 100 steps of the five-state Wigner walk. The initial coin state was chosen as $|\chi_1^-\rangle$. The coin parameter is $\rho=0.3$ }
\label{fig:2:h1m}
\end{figure}

\begin{figure}[h!]
\begin{center}
\includegraphics[width=0.45\textwidth]{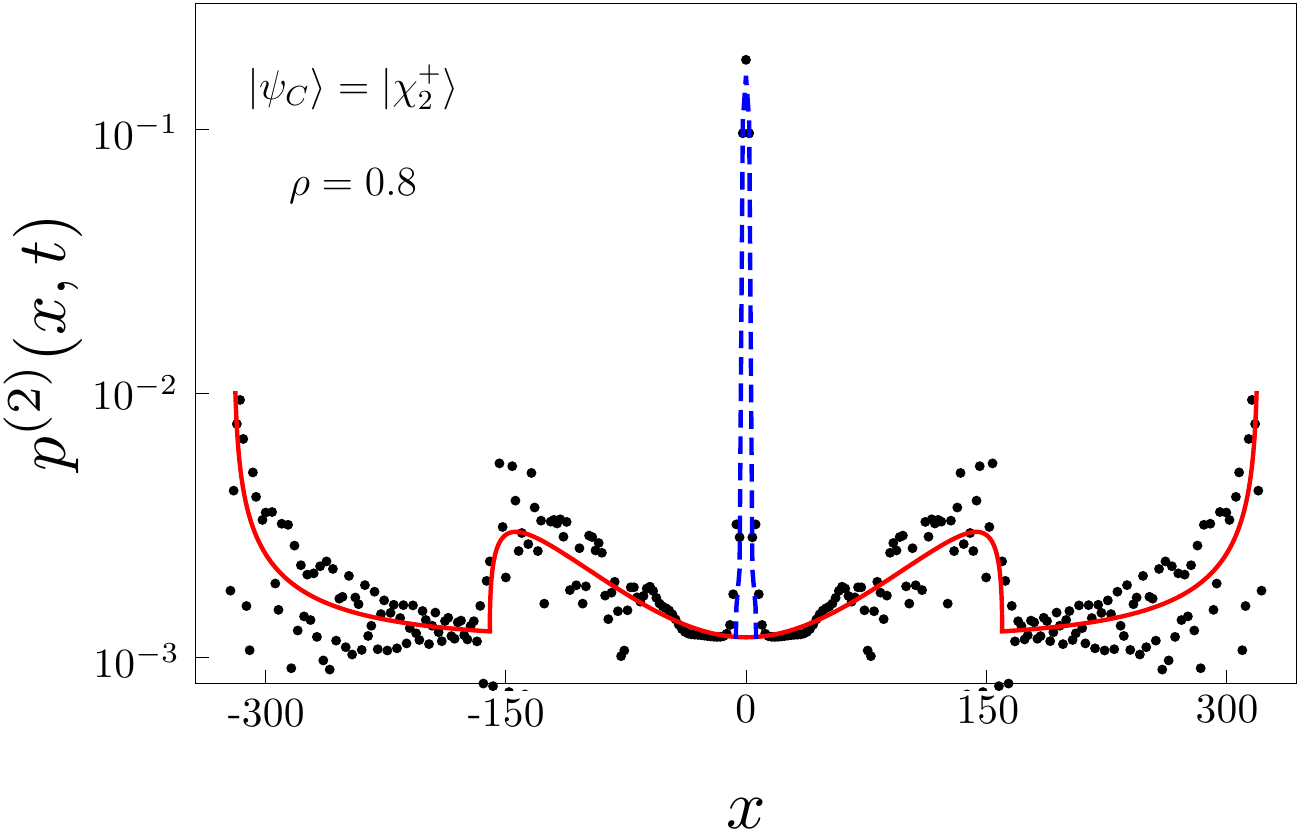}
\end{center}
\caption{(Color
  online) Probability distribution after 100 steps of the five-state Wigner walk. The initial coin state was chosen as $|\chi_2^+\rangle$. The coin parameter is $\rho=0.8$ }
\label{fig:2:h2p}
\end{figure}

\begin{figure}[htbp]
\begin{center}
\includegraphics[width=0.45\textwidth]{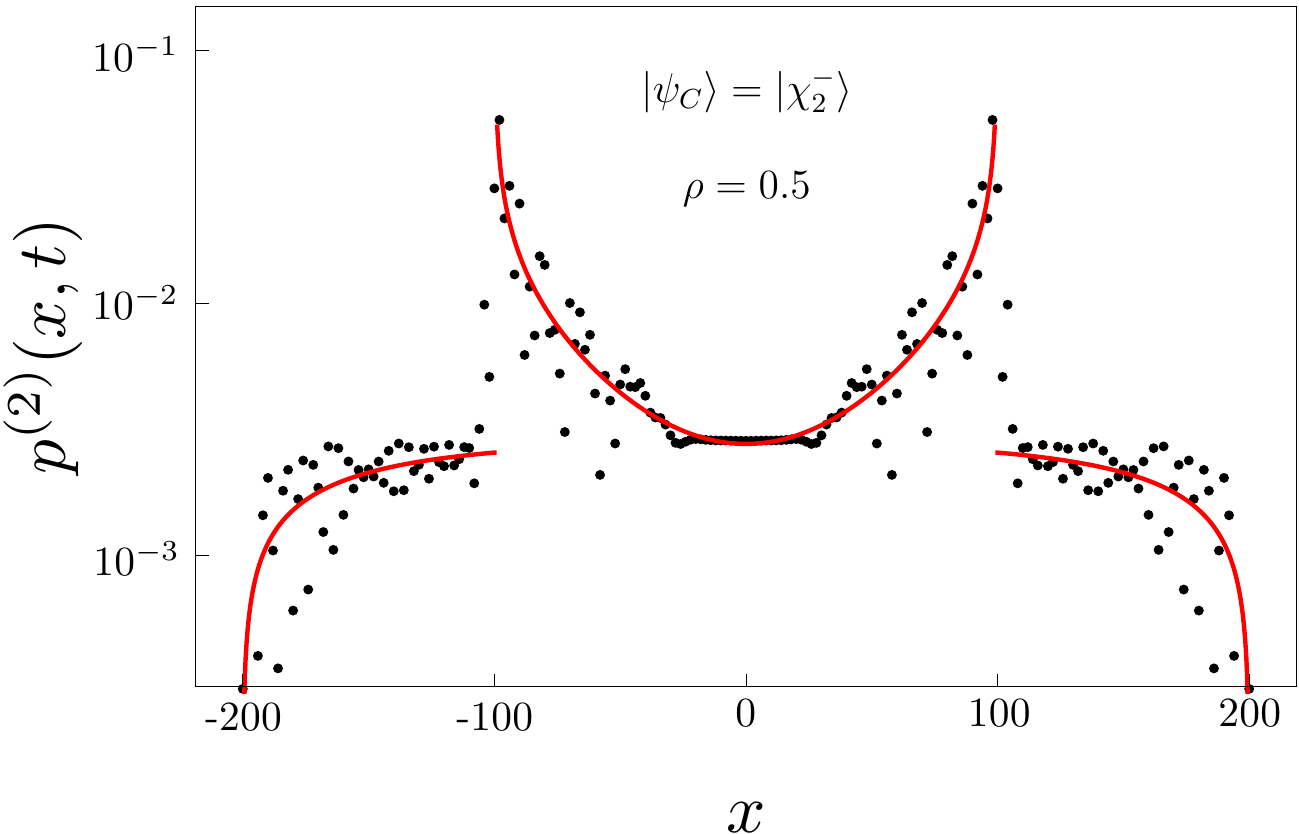}
\end{center}
\caption{(Color
  online) Probability distribution after 100 steps of the five-state Wigner walk. The initial coin state was chosen as $|\chi_2^-\rangle$. The coin parameter is $\rho=0.5$ }
\label{fig:2:h2m}
\end{figure}

In FIG.~\ref{fig:2:h1p} we have chosen the initial state as $|\chi_1^+\rangle$. For this particular state both outer peaks vanishes. FIG.~\ref{fig:2:h1m} displays the probability distribution for the initial state $|\chi_1^-\rangle$. In this situation both inner peaks vanishes and, in addition, the trapping effect disappears. In FIG.~\ref{fig:2:h2p} we have chosen the initial state as $|\chi_2^+\rangle$. For this initial state both inner peaks vanishes. However, unlike in FIG.~\ref{fig:2:h1m} the trapping effect is present. Finally, in FIG.~\ref{fig:2:h2m} we display the probability distribution for the initial state $|\chi_2^-\rangle$. In such a case the outer peaks vanishes. Moreover, the trapping effect is absent, similar to the FIG.~\ref{fig:2:h1m}.

Let us now determine the sets of states for which both peaks of either the slower walk or the faster walk disappear. To achieve this the weight function has to be of the form
$$
{\cal M}^{(2,m)}(v) = (\rho^2-v^2)(a + b v + c v^2),
$$
for some arbitrary $a,b,c$. This is satisfied provided that
\begin{eqnarray}
\label{condPeaks2}
\nonumber {\cal M}^{(2,m)}_1 + \rho^2 {\cal M}^{(2,m)}_3 & = & 0, \\
 {\cal M}^{(2,m)}_0 + \rho^2 {\cal M}^{(2,m)}_2 + \rho^4 {\cal M}^{(2,m)}_4 & = & 0.
\end{eqnarray}
Concerning the slower walk described by the limit density $\nu^{(2,1)}(v)$, from the explicit form of the weight function ${\cal M}^{(2,1)}(v)$ given in (\ref{weight:2:1}) we obtain the conditions
\begin{eqnarray}
\nonumber h_1^+ \overline{h_2^-} + \overline{h_1^+} h_2^- & = & 0,\\
\nonumber  |h_1^+|^2 + |h_2^-|^2 & = & 0,
\end{eqnarray}
which are satisfied for
\begin{equation}
\label{condelimj2m1}
h_1^+ = h_2^- = 0.
\end{equation}
Thus we have a three-dimensional subspace of initial coin states
$$
|\psi_C^{(1)}\rangle  = h_0|\chi_0\rangle  + h_1^-|\chi_1^-\rangle + h_2^+|\chi_2^+\rangle
$$
for which both divergencies in $\nu^{(2,1)}(v)$ disappear.

For the faster walk described by the limit density $\nu^{(2,2)}(v)$ the condition (\ref{condPeaks2}) for the weight ${\cal M}^{(2,2)}(v)$ leads us to the relations
\begin{eqnarray}
\nonumber h_1^-(\sqrt{3}\overline{h_0} - \overline{h_2^+}) + \overline{h_1^-}(\sqrt{3}h_0 - h_2^+)  & = & 0,\\
\nonumber |\sqrt{3}h_0 - h_2^+|^2 + 4|h_1^-|^2 & = & 0.
\end{eqnarray}
This is satisfied provided that
\begin{equation}
\label{condelimj2m2}
 h_1^- = 0,\quad h_2^+  = \sqrt{3}h_0.
\end{equation}
Hence, we find a three-dimensional subspace of initial states
$$
|\psi_C^{(2)}\rangle  = h_0|\chi_0\rangle  + h_1^+|\chi_1^+\rangle + \sqrt{3}h_0|\chi^+\rangle + h_2^-|\chi_2^-\rangle
$$
for which the density $\nu^{(2,2)}(v)$ has no divergencies.

\begin{figure}[htbp]
\begin{center}
\includegraphics[width=0.45\textwidth]{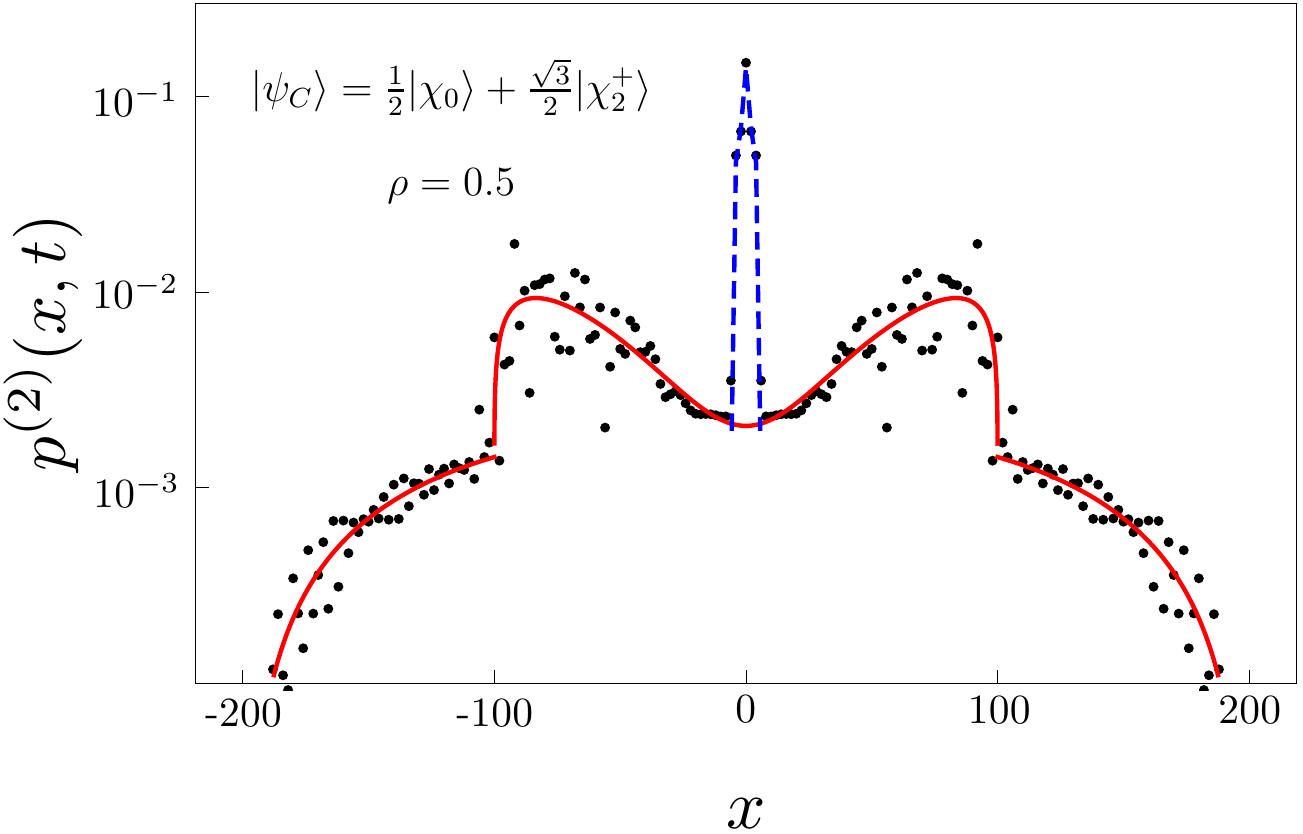}
\end{center}
\caption{(Color
  online) Probability distribution after 100 steps of the five-state Wigner walk. The coin parameter is $\rho=0.5$. The initial coin state was chosen according to (\ref{j=3/2:nodiv}), which satisfies both conditions for eliminations of peaks of the slower walk  (\ref{condelimj2m1}) and the faster walk (\ref{condelimj2m2}) simultaneously. The resulting probability distribution has only one peak at the origin due to the trapping effect.}
\label{fig:nodiv}
\end{figure}

In contrast to the four-state walk which we have treated in the previous Section, it is now possible to satisfy both conditions (\ref{condelimj2m1}) and  (\ref{condelimj2m2}) simultaneously, i.e. we can eliminate all four divergencies in both densities. The state for which this situation occurs is given by
\begin{equation}
|\psi_C\rangle = \frac{1}{2}|\chi_0\rangle + \frac{\sqrt{3}}{2}|\chi_2^+\rangle.
\label{j=3/2:nodiv}
\end{equation}
We point out that this is the only coin state for which the probability distribution of the five-state Wigner walk has only one peak. This peak arises due to the trapping effect which we address in the following subsection. We illustrate this feature in FIG.~\ref{fig:nodiv}.

\begin{figure}[htbp]
\begin{center}
\includegraphics[width=0.45\textwidth]{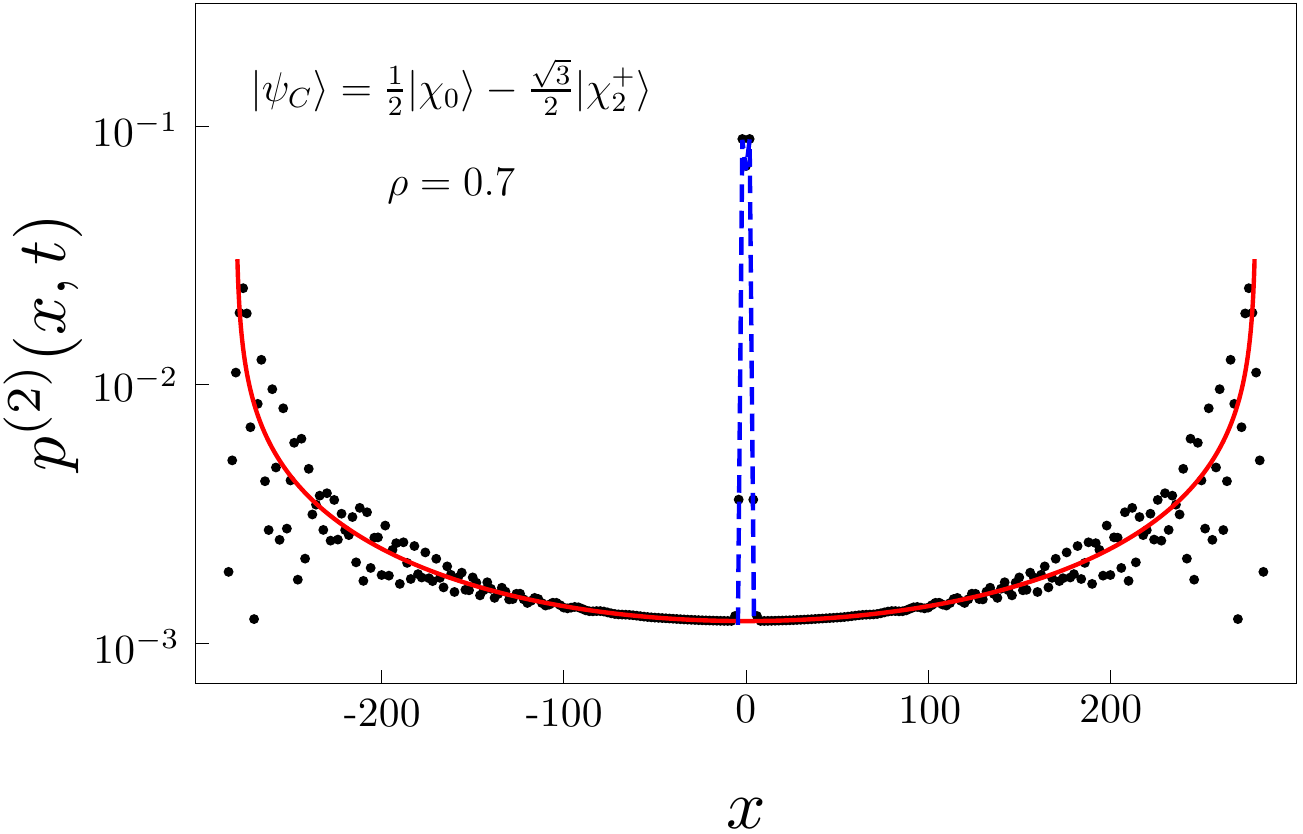}
\end{center}
\caption{(Color
  online) Probability distribution after 100 steps of the five state Wigner walk. The coin parameter is $\rho=0.7$.  The initial state was chosen according to (\ref{j=3/2:noslower}). For this particular initial coin state the limit density $\nu^{(2,1)}(v)$ describing the slower walk vanishes.}
\label{fig:noslower}
\end{figure}

Finally, we note that the slower walk described by the limit density $\nu^{(2,1)}(v)$ can vanish completely. Indeed, for the state
\begin{equation}
|\psi_C\rangle = \frac{1}{2}|\chi_0\rangle - \frac{\sqrt{3}}{2}|\chi_2^+\rangle,
\label{j=3/2:noslower}
\end{equation}
all terms ${\cal M}^{(2,1)}_i$ are equal to zero. In such a case the spreading of the five-state Wigner walk is described only by the limit density $\nu^{(2,2)}(v)$. We illustrate this effect in FIG.~\ref{fig:noslower}.

\subsection{Trapping probability}
\label{sec:2b}

Let us now turn to the trapping probability. We leave the details of calculations for the Appendix~\ref{app:5state} and focus on the dependence of the trapping probability on the initial coin state. As we discuss in more detail in the Appendix~\ref{app:5state}, the trapping effect does not occur when the initial state is a linear combination of vectors $|\chi_1^-\rangle$ and $|\chi_2^-\rangle$. This feature was illustrated already in FIGs.~\ref{fig:2:h1m} and \ref{fig:2:h2m}. It implies that in the suitable basis $\{|\chi_0\rangle,\ |\chi_1^\pm\rangle,\ |\chi_2^\pm\}$ the trapping probability depends only on three amplitudes of the initial coin state, namely $h_0$, $h_1^+$ and $h_2^+$. However, the explicit form of $p_\infty^{(2)}(x)$ is still rather involved. Nevertheless, this can be overcome by an additional change of basis in the subspace spanned by $\{|\chi_0\rangle,\ |\chi_1^+\rangle,\ |\chi_2^+\}$ which affects the trapping effect. Similarly as for the  three-state model, which we have treated in Section~\ref{sec:1}, there are coin states for which the trapping effect appears only for $x\geq 0$, respectively  $x\leq 0$. We find that the state resulting in trapping of the particle only at non-negative positions is given by
$$
|\lambda^+\rangle = \sqrt{\frac{3}{8}}|\chi_0\rangle + \frac{1}{\sqrt{2}}|\chi_1^+\rangle + \frac{1}{\sqrt{8}}|\chi_2^+\rangle.
$$
The second state, which traps the particle only at positions $x\leq 0$ is orthogonal to $|\lambda^+\rangle$ and reads
$$
|\lambda^-\rangle = \sqrt{\frac{3}{8}}|\chi_0\rangle - \frac{1}{\sqrt{2}}|\chi_1^+\rangle + \frac{1}{\sqrt{8}}|\chi_2^+\rangle.
$$
We find that the orthogonal complement of $|\lambda^\pm\rangle$ within the subspace spanned by $\{|\chi_0\rangle,\ |\chi_1^+\rangle,\ |\chi_2^+\}$ is given by the vector (\ref{j=3/2:noslower}), which we now denote as $|\lambda_0\rangle$. The triplet $\{|\lambda_0\rangle,\ |\lambda^\pm\rangle\}$ forms an orthonormal basis in the subspace affecting the trapping effect, and together with the vectors $|\chi_1^-\rangle$ and $|\chi_2^-\rangle$ it forms an orthonormal basis of the whole coin space. When we express the initial state in the new basis according to
$$
|\psi_C\rangle = l_0 |\lambda_0\rangle + l^+|\lambda^+\rangle + l^-|\lambda^-\rangle + h_1^-|\chi_1^- \rangle + h_2^-|\chi_2^- \rangle,
$$
we find that the trapping probability for positive $x$ reads
\begin{equation}
\label{j:2:locp}
p^{(2)}_\infty(2x) = Q^{2x} \frac{3(1-\rho^2)}{2\rho^4}\left(|l_0 + f(x)l^+|^2 + |l^+|^2 \right),
\end{equation}
and similarly for negative $x$ we find
\begin{equation}
\label{j:2:locm}
p^{(2)}_\infty(2x) = Q^{2|x|} \frac{3(1-\rho^2)}{2\rho^4}\left(|l_0 + f(x)l^-|^2 + |l^-|^2 \right).
\end{equation}
Here we have used the notation
\begin{eqnarray}
\nonumber f(x) & = & \frac{\sqrt{6}}{\rho^2}\left(\rho^2 - 2 + 2|x|\sqrt{1-\rho^2}\right).
\end{eqnarray}
Directly at the origin the trapping probability has a more complicated form
\begin{eqnarray}
\label{j:2:loc0}
\nonumber p^{(2)}_\infty(0) & = & \frac{9(1-\rho^2)}{4\rho^4}Q^2\left(|l_+|^2 + |l_-|^2\right) + \\
\nonumber & & + \frac{3}{8}Q^2|l_+ + l_-|^2  + \frac{2-\rho^2-\sqrt{1-\rho^2}}{4 \rho^2} Q |l_0|^2 - \\
\nonumber & & -\frac{\sqrt{6} \left(2 - \rho^2 + \frac{1}{2}\sqrt{1-\rho^2}\right)}{8 \rho^2} Q^2\times \\
& & \times \left((l_+ + l_-)\overline{l_0} + (\overline{l_+} + \overline{l_-})l_0\right).
\end{eqnarray}
The results (\ref{j:2:locp}) and (\ref{j:2:locm}) indicate that the decay of the trapping probability with distance from the origin is not purely exponential like for the three-state Wigner walk (\ref{trapping1}). Nevertheless, the correction to the exponential decay becomes negligible for large $x$.  The state $|\lambda_0\rangle$ is an exception, since for $l^\pm = 0$ the terms involving the position-dependent function $f(x)$ vanish and the behaviour of the trapping probability is exactly exponential. Moreover, the decay rate determined by $Q$ is the same as for the three-state Wigner walk.

\begin{figure}[htbp]
\begin{center}
\includegraphics[width=0.45\textwidth]{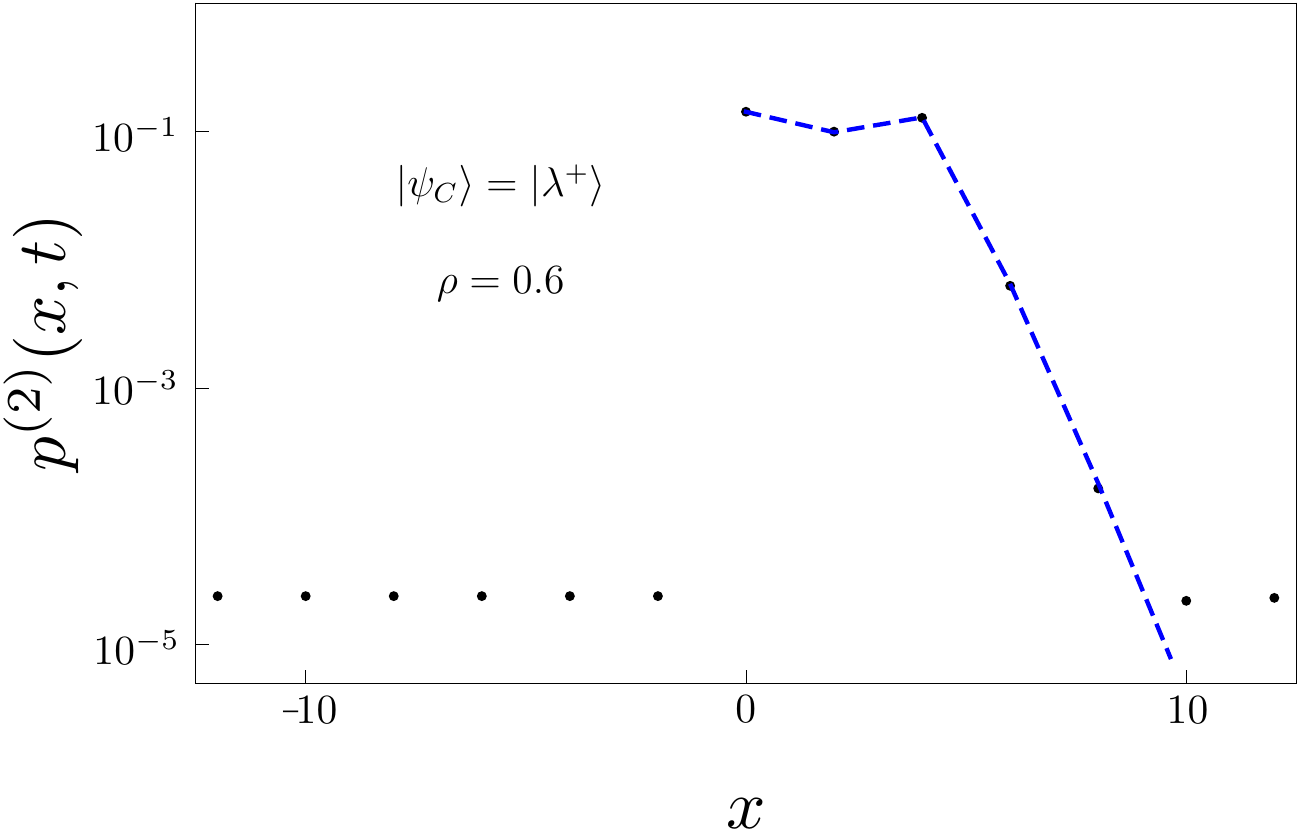}
\end{center}
\caption{(Color
  online) Probability distribution after 10000 steps of the five-state Wigner walk. The initial state was chosen as $|\lambda^+\rangle$. The coin parameter is $\rho=0.6$. Only a small vicinity of the origin is displayed to emphasize that for $|\lambda^+\rangle$ the trapping effect appears only for $x\geq 0$. Higher number of steps in comparison with other figures was chosen so that the trapping effect is sufficiently pronounced further from the origin. Notice the plateau formed by the first three points exemplifying the fact that for $|\lambda^+\rangle$ the decay of the trapping probability (\ref{j:2:locp}) is not purely exponential.}
\label{fig:2:locp}
\end{figure}

For illustration, we display in FIG.~\ref{fig:2:locp} the probability distribution of the five-state Wigner walk with the initial coin state $|\lambda^+\rangle$. Clearly, the trapping effect appears only for $x\geq 0$. Moreover, the plateau formed by the first three point indicates that the decay of the trapping probability deviates from a pure exponential, in accordance with (\ref{j:2:locp}).

\section{Conclusions}
\label{sec:outlook}

We have investigated in detail quantum walk models on a line with coin operators given by Wigner rotation matrices $\hat{R}^{(j)}(\alpha,\beta,\gamma)$. We have shown that this three-parameter set of walks is equivalent to a single parameter family $\hat{R}^{(j)}(\rho)$, as far as the shape of the probability distribution and its moments are concerned. The parameter $\rho$ has a simple physical interpretation since it determines the position of the peaks in the probability distribution and its width.

Next, we have simplified the results of \cite{konno:wigner} by turning to a more suitable basis of the coin space. Unlike for the three-state Grover walk \cite{stef:limit}, the suitable basis is not directly given by the eigenvectors of the coin operator, however, we have found a recipe which allowed us to construct the suitable basis from them. We presented the explicit form of the limit densities for Wigner walks up to $j=2$. Expressing them in the suitable basis allowed us to identify various interesting regimes which are otherwise hidden in the standard basis description. As an example, we have shown that the number of peaks in the probability distribution can be reduced to one. Moreover, the suitable basis reveals that $\rho$ is a scaling parameter, since varying the value of $\rho$ while keeping the same amplitudes of the initial state in the suitable basis simply changes the width of the probability distribution without distorting its shape.

The Wigner walks with integer $j$ show the trapping effect, which was not explicitly evaluated in \cite{konno:wigner}. We presented the explicit form of the trapping probability for $j=1$ and $j=2$. Our results show that the trapping probability can be highly asymmetric and, moreover, is not purely exponential for $j=2$.

For clarity the Tables \ref{tab:3s}, \ref{tab:4s} and \ref{tab:5s} summarize the main effects and its graphical representation in the text for three-, four- and five-state Wigner walks.

\begin{table}
\begin{center}
\begin{tabular}{|c|c|c|}
  \hline
  \multirow{2}{*}{Initial state} & \multirow{2}{*}{Properties} & \multirow{2}{*}{Figure} \\
  & & \\\hline
  \multirow{2}{*}{$|\chi_0\rangle$} & \multirow{2}{*}{density tends to zero at the origin} &  \multirow{2}{*}{FIG.~\ref{fig:1:h0}} \\
  & & \\ \hline
  \multirow{2}{*}{$|\chi^+\rangle$} & \multirow{2}{*}{only one peak due to the trapping} & \multirow{2}{*}{FIG.~\ref{fig:1:hp}} \\
  & & \\ \hline
  \multirow{2}{*}{$|\chi^-\rangle$} & \multirow{2}{*}{no trapping effect} & \multirow{2}{*}{FIG.~\ref{fig:1:hm}} \\
  & & \\ \hline
  \multirow{2}{*}{$|\lambda^-\rangle$} & \multirow{2}{*}{no trapping on the positive half-line} & \multirow{2}{*}{FIG.~\ref{fig:1:lm}}\\
  & & \\ \hline
\end{tabular}
\caption{Summary of figures for the three-state model.}
\label{tab:3s}
\end{center}
\end{table}

\begin{table}
\begin{center}
\begin{tabular}{|c|c|c|}
  \hline
  \multirow{2}{*}{Initial state} & \multirow{2}{*}{Properties} & \multirow{2}{*}{Figure} \\
  & & \\\hline
  \multirow{3}{*}{$|\chi_1^+\rangle$} & \multirow{2}{*}{only one peak in the slower walk} &  \multirow{3}{*}{FIG.~\ref{fig:32:h1}} \\
      & \multirow{2}{*}{only one peak in the faster walk} & \\
  & & \\ \hline
  \multirow{3}{*}{$|\chi_2^+\rangle$} & \multirow{2}{*}{only one peak in the slower walk} & \multirow{3}{*}{FIG.~\ref{fig:32:h2}} \\
        & \multirow{2}{*}{only one peak in the faster walk} & \\
  & & \\ \hline
  \multirow{3}{*}{$\sqrt{3}|\chi_1^-\rangle-|\chi_2^-\rangle$} & \multirow{2}{*}{no peaks in the slower walk} & \multirow{3}{*}{FIG.~\ref{fig:inner:peak}} \\
  & \multirow{2}{*}{only one peak in the faster walk} & \\
  & & \\ \hline
  \multirow{3}{*}{$|\chi_1^+\rangle+ \sqrt{3}|\chi_2^+\rangle$} & \multirow{2}{*}{no peaks in the faster walk} & \multirow{3}{*}{FIG.~\ref{fig:outer:peak}} \\
    & \multirow{2}{*}{only one peak in the slower walk} & \\
  & & \\ \hline
\end{tabular}
\caption{Summary of figures for the four-state model.}
\label{tab:4s}
\end{center}
\end{table}

\begin{table}
\begin{center}
\begin{tabular}{|c|c|c|}
  \hline
  \multirow{2}{*}{Initial state} & \multirow{2}{*}{Properties} & \multirow{2}{*}{Figure} \\
  & & \\\hline
  \multirow{3}{*}{$|\chi_0\rangle$} & \multirow{2}{*}{no peaks in the slower walk} &  \multirow{3}{*}{FIG.~\ref{fig:2:h0}} \\
  & \multirow{2}{*}{density tends to zero at the origin}  & \\
    & & \\ \hline
  \multirow{2}{*}{$|\chi_1^+\rangle$} & \multirow{2}{*}{no peaks in the faster walk} & \multirow{2}{*}{FIG.~\ref{fig:2:h1p}} \\
  & & \\ \hline
  \multirow{3}{*}{$|\chi_1^-\rangle$} & \multirow{2}{*}{no peaks in the slower walk} & \multirow{3}{*}{FIG.~\ref{fig:2:h1m}} \\
    & \multirow{2}{*}{no trapping effect}  & \\
  & & \\ \hline
  \multirow{2}{*}{$|\chi_2^+\rangle$} & \multirow{2}{*}{no peaks in the slower walk} & \multirow{2}{*}{FIG.~\ref{fig:2:h2p}} \\
  & & \\ \hline
  \multirow{3}{*}{$|\chi_2^-\rangle$} & \multirow{2}{*}{no peaks in the faster walk} & \multirow{3}{*}{FIG.~\ref{fig:2:h2m}} \\
      & \multirow{2}{*}{no trapping effect}  & \\
    & & \\ \hline
  \multirow{2}{*}{$|\chi_0\rangle + \sqrt{3}|\chi_2^+\rangle$} & \multirow{2}{*}{only one peak due to the trapping} & \multirow{2}{*}{FIG.~\ref{fig:nodiv}} \\
  & & \\ \hline
  \multirow{2}{*}{$|\chi_0\rangle - \sqrt{3}|\chi_2^+\rangle$} & \multirow{2}{*}{slower walk vanishes} & \multirow{2}{*}{FIG.~\ref{fig:noslower}} \\
  & & \\ \hline
  \multirow{2}{*}{$|\lambda^+\rangle$} & \multirow{2}{*}{no trapping on the negative half-line} & \multirow{2}{*}{FIG.~\ref{fig:2:locp}} \\
  & & \\ \hline
\end{tabular}
\caption{Summary of figures for the five-state model.}
\label{tab:5s}
\end{center}
\end{table}

Based on our explicit results for Wigner walks up to $j=2$ we make the following conjecture on the suitable basis for models with greater $j$. For half-integer $j$ the coin operator $\hat{R}^{(j)}(\rho)$ has $\left\lfloor\frac{2j+1}{2}\right\rfloor$ pairs of eigenvectors $|\psi_n^{\pm}\rangle$ corresponding to eigenvalues of the form $e^{\pm i\varphi_n}$.  The suitable basis $\{|\chi_n^{\pm}\rangle\}$ can be constructed according to the recipe
\begin{eqnarray}
\nonumber |\chi_n^+\rangle & = & \frac{1}{\sqrt{2}}\left(e^{-i\frac{\varphi}{2}}|\psi_n^+\rangle + e^{i\frac{\varphi}{2}}|\psi_n^-\rangle\right),\\
\nonumber |\chi_n^-\rangle & = & \frac{i}{\sqrt{2}}\left(e^{-i\frac{\varphi}{2}}|\psi_n^+\rangle - e^{i\frac{\varphi}{2}}|\psi_n^-\rangle\right).
\end{eqnarray}
For integer $j$ there is an additional eigenvector $|\psi_0\rangle \equiv |\chi_0\rangle$ with eigenvalue 1, which we include as the last vector of the suitable basis.

The vectors of the suitable basis were selected among those which result in non-generic probability distribution of the Wigner walk. They point out the extremal regimes in which these quantum walks can be operated. Moreover, they indicate ways in which wave-packet could be shaped using quantum walks with higher-dimensional coins. We assume that this approach will be useful in analyzing the properties of other quantum walk models, especially on higher dimensional lattices.

\section*{Acknowledgement}

M\v S is grateful for the financial support from GA\v CR 14-02901P. IB and IJ are grateful for the financial support from SGS13/217/OHK4/3T/14 and GA\v CR 13-33906S.

\appendix

\section{Three-state model}
\label{app:3state}

\subsection{Suitable basis}

For the three-state model the explicit form of the eigenvectors of the coin ${\hat R}^{(1)}(\rho)$ reads
\begin{eqnarray}
\label{app:evec:1}
\nonumber |\psi^\pm\rangle & = & \frac{1}{2}(\pm i|1\rangle + \sqrt{2}|0\rangle \mp i|-1\rangle),\\
|\psi_0\rangle & = & \frac{1}{\sqrt{2}}(|1\rangle + |-1\rangle).
\end{eqnarray}
The eigenvectors satisfy the equations
\begin{eqnarray}
\nonumber {\hat R}^{(1)}(\rho)|\psi^\pm\rangle & = & e^{\pm i\varphi}|\psi^\pm\rangle,\\
\nonumber {\hat R}^{(1)}(\rho)|\psi_0\rangle & = & |\psi_0\rangle,
\end{eqnarray}
where the phase $\varphi$ is given by
$$
\varphi = \arccos{(2\rho^2-1)}.
$$
The suitable basis is then constructed according to the formula (\ref{basis:1}). We find that the vectors of the suitable basis $\{|\chi_0\rangle, |\chi^\pm\rangle \}$ are in terms of the standard basis given by
\begin{eqnarray}
\nonumber |\chi_0\rangle & = & \frac{1}{\sqrt{2}}(|1\rangle + |-1\rangle),\\
\nonumber |\chi^+\rangle & = & \sqrt{\frac{1-\rho^2}{2}}|1\rangle + \rho |0\rangle -\sqrt{\frac{1-\rho^2}{2}} |-1\rangle,\\
\nonumber |\chi^-\rangle & = & -\frac{\rho}{\sqrt{2}}|1\rangle +\sqrt{1-\rho^2}|0\rangle +\frac{\rho}{\sqrt{2}} |-1\rangle.
\end{eqnarray}
This leads us to the following relations between the coefficients of the initial state when expressed in the standard basis ($q_i$) and the suitable basis ($h_i$)
\begin{eqnarray}
\label{app:sb:1}
\nonumber q_1 & = & \frac{1}{\sqrt{2}}h_0 + \sqrt{\frac{1-\rho^2}{2}}h^+ - \frac{\rho}{\sqrt{2}} h^-,\\
\nonumber q_0 & = & \rho h^+ + \sqrt{1-\rho^2}h^-,\\
q_{-1} & = & \frac{1}{\sqrt{2}}h_0 - \sqrt{\frac{1-\rho^2}{2}}h^+ + \frac{\rho}{\sqrt{2}} h^-.
\end{eqnarray}

\subsection{Weight function}

The individual terms of the weight ${\cal M}^{(1,1)}(v)$ are in the standard basis of the coin space given by \cite{konno:wigner}
\begin{eqnarray}
\nonumber {\cal M}_0^{(1,1)} & = & \frac{1}{2}\left(|q_1|^2 + 2 |q_0|^2 + |q_{-1}|^2 - q_1\overline{q_{-1}} - \overline{q_1}q_{-1}\right),\\
\nonumber {\cal M}_1^{(1,1)} & = & -|q_1|^2 + |q_{-1}|^2 + \\
\nonumber & & + \frac{\sqrt{1-\rho^2}}{\sqrt{2}\rho}(q_1\overline{q_0} + \overline{q_1}q_0 + q_0\overline{q_{-1}} + \overline{q_0}q_{-1}),\\
\nonumber {\cal M}_2^{(1,1)} & = & \frac{1}{2}(|q_1|^2 - 2|q_0|^2 + |q_{-1}|^2) - \\
\nonumber & & - \frac{\sqrt{1-\rho^2}}{\sqrt{2}\rho}(q_1\overline{q_0} + \overline{q_1}q_0 - q_0\overline{q_{-1}} - \overline{q_0}q_{-1}) + \\
\nonumber  & &  + \frac{2-\rho^2}{2\rho^2}(q_1\overline{q_{-1}} + \overline{q_1}q_{-1} ).
\end{eqnarray}
Using the relations (\ref{app:sb:1}) we find that in the suitable basis $\{|\chi_0\rangle, |\chi^\pm\rangle \}$ the weight function simplifies into
\begin{eqnarray}
\nonumber {\cal M}_0^{(1,1)} & = & |h^+|^2 + |h^-|^2,\\
\nonumber {\cal M}_1^{(1,1)} & = & \frac{1}{\rho}(h_0\overline{h^-} + \overline{h_0} h^-),\\
\nonumber {\cal M}_2^{(1,1)} & = & \frac{1}{\rho^2}(|h_0|^2 - |h^+|^2).
\end{eqnarray}

\subsection{Trapping probability}

The trapping effect arises from the fact that the evolution operator of the three-state Wigner walk has an infinitely degenerate eigenvalue 1. The overlap of the corresponding localized eigenvectors with the initial state leads to the non-vanishing trapping probability (\ref{trap:prob}). The overlap is simple to evaluate in the Fourier picture \cite{ambainis}. Indeed, the Fourier transformation diagonalizes the step operator (\ref{step}) and the evolution operator (\ref{evol:U}) reduces into a $3\times 3$ matrix
$$
\tilde{U}(k) = {\rm Diag}(e^{2i k},\ 1,\ e^{-2i k})\cdot R^{(1)}(\rho),
$$
where $k$ is a continuous quasi-momentum ranging from $0$ to $2\pi$. The matrix $\tilde{U}(k)$ has an $k$-independent eigenvalue 1 with the eigenvector
\begin{equation}
v(k) = \frac{1}{\sqrt{4 - 2\rho^2(1+\cos{2k})}}\left(
                                                                   \begin{array}{c}
                                                                     \sqrt{2} \sqrt{1-\rho^2} \\
                                                                     \left(1-e^{2 i k}\right) \rho \\
                                                                     \sqrt{2} e^{2 i k} \sqrt{1-\rho^2} \\
                                                                   \end{array}
                                                                 \right).
\label{stat:1}
\end{equation}
The overlap of the eigenvector with the Fourier transformed initial state
\begin{equation}
\label{app:init:FT}
\tilde{\psi}_C = \left(q_{1}, q_0, q_{-1}\right)^T,
\end{equation}
where $q_i$ are expressed in (\ref{app:sb:1}), yields the limiting probability amplitude at position $x$
\begin{equation}
\label{limit:ampl}
\psi_\infty^{(1)}(x) = \int\limits_{0}^{2\pi} \frac{dk}{2\pi} e^{- i x k} (v(k),\tilde{\psi}_C) v(k).
\end{equation}
Direct evaluation of the scalar product $(v(k),\tilde{\psi}_C)$ reveals that it vanishes when the initial coin state is $|\chi^-\rangle$. Hence, for this particular state the trapping effect does not occur. For other initial states the limiting amplitude can be decomposed into integrals of the form
$$
{\cal I}^{(1)}(x) = \int\limits_{0}^{2\pi} \frac{e^{-i x k}}{4\pi(2-\rho^2(1+\cos{k}))} dk.
$$
Using the substitution $z = e^{i k}$ one can turn ${\cal I}^{(1)}(x)$ to a contour integral over a unit circle in the complex plane, which can be evaluated explicitly by the method of residues. We find that the result reads
$$
{\cal I}^{(1)}(x) = \frac{Q^{|x|}}{4\sqrt{1-\rho^2}},
$$
where the factor $Q$ is given in (\ref{Q}). Finally, the trapping probability $p_\infty^{(1)}(x)$ is given by the square norm of the amplitude (\ref{limit:ampl}). The final result is presented in (\ref{trapping1})

\section{Four-state model}
\label{app:4state}

\subsection{Suitable basis}
\label{app:3/2:1}

For the four-state model the coin operator ${\hat R}^{(3/2)}(\rho)$ has two pairs of eigenvectors
\begin{eqnarray}
\label{app:evec:3/2}
\nonumber |\psi_1^\pm\rangle & = & \frac{1}{\sqrt{8}}(\sqrt{3}|3/2\rangle \mp i|1/2\rangle + |-1/2\rangle \mp i\sqrt{3}|-3/2\rangle),\\
\nonumber |\psi_2^\pm\rangle & = & \frac{1}{\sqrt{8}}(|3/2\rangle \pm i \sqrt{3}|1/2\rangle -\sqrt{3} |-1/2\rangle  \mp i|-3/2\rangle),\\
\end{eqnarray}
with conjugated pairs of eigenvalues
\begin{eqnarray}
\nonumber {\hat R}^{(3/2)}(\rho)|\psi_1^\pm\rangle & = & e^{\pm i\varphi_1}|\psi_1^\pm\rangle,\\
\nonumber {\hat R}^{(3/2)}(\rho)|\psi_2^\pm\rangle & = & e^{\pm i\varphi_2}|\psi_2^\pm\rangle.
\end{eqnarray}
The phases of the eigenvalues are given by
\begin{eqnarray}
\nonumber \varphi_1 & = & \arccos{\rho},\\
\nonumber \varphi_2 & = & \left\{
                            \begin{array}{c}
                              \arccos\left(\rho(4\rho^2-3)\right),\quad 0<\rho\leq \frac{1}{2} \\
                               \\
                              2\pi - \arccos\left(\rho(4\rho^2-3)\right),\quad \frac{1}{2}<\rho\leq 1 \\
                            \end{array}
                          \right. .
\end{eqnarray}
The suitable basis is then constructed according to (\ref{basis:3:2}). Using the explicit form of the eigenvectors (\ref{app:evec:3/2}) we can obtain direct relation between the standard basis $\{|3/2\rangle,|1/2\rangle,|-1/2\rangle,|-3/2\rangle\}$ and the suitable basis $\{|\chi_1^\pm\rangle,|\chi_2^\pm\rangle\}$. From this we find the following correspondence between the coefficients of the initial state in the standard basis ($q_i$) and in the suitable basis ($h_i$)
\begin{widetext}

\begin{eqnarray}
\label{app:sb:3/2}
\nonumber q_{3/2} & = & \frac{\sqrt{3(1+\rho)}}{2 \sqrt{2}}h_1^+ + \frac{\sqrt{3(1-\rho)}}{2 \sqrt{2}}h_1^- + \frac{\sqrt{1+\rho}}{2 \sqrt{2}}(1-2\rho) h_2^+  + \frac{\sqrt{1-\rho}}{2 \sqrt{2}}(1+2\rho) h_2^-,\\
\nonumber q_{1/2} & = & - \frac{\sqrt{1-\rho}}{2 \sqrt{2}}h_1^+ + \frac{\sqrt{1+\rho}}{2 \sqrt{2}}h_1^- + \frac{\sqrt{3(1-\rho)}}{2 \sqrt{2}}(1+2\rho) h_2^+  - \frac{\sqrt{3(1+\rho)}}{2 \sqrt{2}}(1-2\rho) h_2^-,\\
\nonumber q_{-1/2} & = & \frac{\sqrt{1+\rho}}{2 \sqrt{2}}h_1^+ + \frac{\sqrt{1-\rho}}{2 \sqrt{2}}h_1^- - \frac{\sqrt{3(1+\rho)}}{2 \sqrt{2}}(1-2\rho) h_2^+  - \frac{\sqrt{3(1-\rho)}}{2 \sqrt{2}}(1+2\rho) h_2^-,\\
q_{-3/2} & = & -\frac{\sqrt{3(1-\rho)}}{2 \sqrt{2}}h_1^+ + \frac{\sqrt{3(1+\rho)}}{2 \sqrt{2}}h_1^- -\frac{\sqrt{1-\rho}}{2 \sqrt{2}}(1+2\rho) h_2^+ + \frac{\sqrt{1+\rho}}{2 \sqrt{2}}(1-2\rho) h_2^-.
\end{eqnarray}

\subsection{Weight function}
\label{app:3/2:2}

The weight functions ${\cal M}^{(3/2,1/2)}(v)$ and ${\cal M}^{(3/2,3/2)}(v)$ for the four-state model in terms of the standard basis coefficients $q_i$ were given explicitly in \cite{konno:wigner}, we do not reproduce it since it is rather long. With the help of the relations (\ref{app:sb:3/2}) we can express them in terms of the suitable basis amplitudes $h_i$. We find that the individual terms of the weight function ${\cal M}^{(3/2,1/2)}(v)$ obtain the form
\begin{eqnarray}
\label{weight:3/2:1/2}
\nonumber  {\cal M}_0^{(3/2,1/2)} & = & |h_1^+|^2 + |h_1^-|^2, \\
\nonumber {\cal M}_1^{(3/2,1/2)} & = & -\frac{1}{\rho} \left[2(|h_1^+|^2-|h_1^-|^2) + \frac{\sqrt{3}}{2}(h_1^+\overline{h_2^+} + \overline{h_1^+} h_2^+ - h_1^-\overline{h_2^-} - \overline{h_1^-} h_2^-)\right],\\
\nonumber {\cal M}_2^{(3/2,1/2)} & = & -\frac{\sqrt{3}}{4\rho^2}\left[\sqrt{3}(|h_1^+|^2  +|h_1^-|^2 - |h_2^+|^2 - |h_2^-|^2) - (h_1^+\overline{h_2^+} + \overline{h_1^+} h_2^+ + h_1^-\overline{h_2^-} + \overline{h_1^-} h_2^-)\right],\\
{\cal M}_3^{(3/2,1/2)} & = & \frac{3}{4\rho^3}\left[(3|h_1^+|^2 - 3|h_1^-|^2 + |h_2^+|^2 - |h_2^-|^2) + \sqrt{3}(h_1^+\overline{h_2^+} + \overline{h_1^+} h_2^+ - h_1^-\overline{h_2^-} - \overline{h_1^-} h_2^-)\right] .
\end{eqnarray}
Similarly, the coefficients of the weight function for ${\cal M}^{(3/2,3/2)}(v)$ read
\begin{eqnarray}
\label{weight:3/2:3/2}
\nonumber {\cal M}_0^{(3/2,3/2)} & = & |h_2^+|^2 + |h_2^-|^2,\\
\nonumber {\cal M}_1^{(3/2,3/2)} & = & \frac{\sqrt{3}}{2\rho}(h_1^+\overline{h_2^+} + \overline{h_1^+} h_2^+ - h_1^-\overline{h_2^-} - \overline{h_1^-} h_2^-), \\
\nonumber {\cal M}_2^{(3/2,3/2)} & = & \frac{\sqrt{3}}{4\rho^2}\left[\sqrt{3}(|h_1^+|^2 +|h_1^-|^2-|h_2^+|^2-|h_2^-|^2) - (h_1^+\overline{h_2^+} + \overline{h_1^+} h_2^+ + h_1^-\overline{h_2^-} + \overline{h_1^-} h_2^-)\right],\\
{\cal M}_3^{(3/2,3/2)} & = & -\frac{1}{4\rho^3}\left[(3|h_1^+|^2-3|h_1^-|^2+|h_2^+|^2-|h_2^-|^2) + \sqrt{3}(h_1^+\overline{h_2^+} + \overline{h_1^+} h_2^+ - h_1^-\overline{h_2^-} - \overline{h_1^-} h_2^-)\right].
\end{eqnarray}

\end{widetext}

\section{Five-state model}
\label{app:5state}

\subsection{Suitable basis}

The eigenvectors of the coin operator ${\hat R}^{(2)}(\rho)$ for  the five-state model are given by
\begin{eqnarray}
\label{app:evec:2}
\nonumber |\psi_1^\pm\rangle & = & \frac{1}{2}(\pm i|2\rangle + |1\rangle + |-1\rangle \mp i|-2\rangle),\\
\nonumber |\psi_2^\pm\rangle & = & \frac{1}{4}(|2\rangle \pm 2i|1\rangle -\sqrt{6}|0\rangle \mp 2i|-1\rangle + |-2\rangle),\\
|\psi_0\rangle & = & \sqrt{\frac{3}{8}} |2\rangle + \frac{1}{2}|0\rangle + \sqrt{\frac{3}{8}} |-2\rangle.
\end{eqnarray}
They satisfy the eigenvalue equations
\begin{eqnarray}
\nonumber {\hat R}^{(2)}(\rho)|\psi_1^\pm\rangle & = & e^{\pm i \varphi_1}|\psi_1^\pm\rangle, \\
\nonumber {\hat R}^{(2)}(\rho)|\psi_2^\pm\rangle & = & e^{\pm i \varphi_2}|\psi_2^\pm\rangle, \\
\nonumber {\hat R}^{(2)}(\rho)|\psi_0\rangle & = & |\psi_0\rangle,
\end{eqnarray}
with phases determined by
\begin{eqnarray}
\nonumber \varphi_1 & = & \arccos\left(2\rho^2-1\right),\\
\nonumber \varphi_2 & = & \left\{
                            \begin{array}{c}
                              \arccos\left(8\rho^4- 8\rho^2 +1\right),\quad 0<\rho\leq\frac{1}{\sqrt{2}} \\
                               \\
                              2\pi-\arccos\left(8\rho^4- 8\rho^2 +1\right),\quad \frac{1}{\sqrt{2}}<\rho<1 \\
                            \end{array}
                          \right. .
\end{eqnarray}
The suitable basis $\{|\chi_0\rangle,|\chi_1^\pm\rangle,|\chi_2^\pm\rangle\}$ for the five-state model is then constructed according to the formula (\ref{basis:2}). Using the explicit form of the eigenvectors (\ref{app:evec:2}) we can obtain the relation between the vectors of the standard basis and the suitable basis, which leads us to the correspondence between the coefficients of the initial state in the standard basis and the suitable basis
\begin{widetext}
\begin{eqnarray}
\label{app:sb:2}
\nonumber q_2 & = & \frac{\sqrt{3}}{2\sqrt{2}}h_0 + \sqrt{\frac{1-\rho^2}{2}} h_1^+ - \frac{\rho}{\sqrt{2}} h_1^- + \frac{1-2\rho^2}{2\sqrt{2}}h_2^+ + \rho\sqrt{\frac{1-\rho^2}{2}}h_2^-,\\
\nonumber q_1 & = & \frac{\rho}{\sqrt{2}}h_1^+ + \sqrt{\frac{1-\rho^2}{2}}h_1^- + \rho\sqrt{2(1-\rho^2)}h_2^+ - \frac{1-2\rho^2}{\sqrt{2}}h_2^-,\\
\nonumber q_0 & = & \frac{1}{2} h_0 - \frac{\sqrt{3}}{2}(1-2\rho^2)h_2^+ - \rho\sqrt{3(1-\rho^2)}h_2^-,\\
\nonumber q_{-1} & = & \frac{\rho}{\sqrt{2}}h_1^+ + \sqrt{\frac{1-\rho^2}{2}}h_1^- - \rho\sqrt{2(1-\rho^2)}h_2^+ + \frac{1-2\rho^2}{\sqrt{2}}h_2^-,\\
q_{-2} & = & \frac{\sqrt{3}}{2\sqrt{2}}h_0 - \sqrt{\frac{1-\rho^2}{2}} h_1^+ + \frac{\rho}{\sqrt{2}} h_1^- + \frac{1-2\rho^2}{2\sqrt{2}}h_2^+ + \rho\sqrt{\frac{1-\rho^2}{2}}h_2^-.
\end{eqnarray}

\subsection{Weight function}

The weight functions ${\cal M}^{(2,1)}(v)$ and ${\cal M}^{(2,2)}(v)$ for the five-state model in the standard basis of the coin space can be evaluated using the procedure of \cite{konno:wigner}. Using the relations (\ref{app:sb:2}) we can transform them into the suitable basis. We find that the coefficients of the weight function ${\cal M}^{(2,1)}(v)$ obtain the form
\begin{eqnarray}
\label{weight:2:1}
\nonumber {\cal M}^{(2,1)}_0 & = & |h_1^+|^2 + |h_1^-|^2,\\
\nonumber {\cal M}^{(2,1)}_1 & = & \frac{1}{\rho}\left[h_1^+\overline{h_2^-} + \overline{h_1^+}h_2^- + h_2^+ \overline{h_1^-} + \overline{h_2^+} h_1^- + \sqrt{3}(h_0\overline{h_1^-} + \overline{h_0}h_1^-)\right],\\
\nonumber {\cal M}^{(2,1)}_2 & = & \frac{1}{\rho^2}\left[3|h_0|^2 - 4|h_1^+|^2 - |h_1^-|^2  + |h_2^+|^2 + |h_2^-|^2 + \sqrt{3}(h_0\overline{h_2^+} + \overline{h_0}h_2^+)\right],\\
\nonumber {\cal M}^{(2,1)}_3 & = & -\frac{1}{\rho^3}\left[2(h_1^+\overline{h_2^-} + \overline{h_1^+}h_2^-) + h_2^+\overline{h_1^-} + \overline{h_2^+}h_1^- + \sqrt{3}(h_0\overline{h_1^-} + \overline{h_0}h_1^-)\right],\\
{\cal M}^{(2,1)}_4 & = & -\frac{1}{\rho^4}\left[3|h_0|^2 - 4|h_1^+|^2 + |h_2^+|^2 + \sqrt{3}(h_0\overline{h_2^+} + \overline{h_0}h_2^+)\right].
\end{eqnarray}
Similarly, the coefficients of the weight function ${\cal M}^{(2,2)}(v)$ in the suitable basis are easily found to be
\begin{eqnarray}
\label{weight:2:2}
\nonumber {\cal M}^{(2,2)}_0 & = & |h_2^+|^2 + |h_2^-|^2, \\
\nonumber {\cal M}^{(2,2)}_1 & = & - \frac{1}{\rho}\left[h_1^+\overline{h_2^-} + \overline{h_1^+}h_2^- + h_1^-\overline{h_2^+} + \overline{h_1^-}h_2^+\right], \\
\nonumber {\cal M}^{(2,2)}_2 & = & \frac{1}{\rho^2}\left[|h_1^+|^2 + |h_1^-|^2 - |h_2^+|^2 - |h_2^-|^2 - \frac{\sqrt{3}}{2}(h_0\overline{h_2^+} + \overline{h_0}h_2^+)\right], \\
\nonumber {\cal M}^{(2,2)}_3 & = &  \frac{1}{2\rho^3}\left[2(h_1^+\overline{h_2^-} + \overline{h_1^+}h_2^-)  + (h_1^-\overline{h_2^+} + \overline{h_1^-}h_2^+) + \sqrt{3}(h_0\overline{h_1^-} + \overline{h_0}h_1^-)\right], \\
{\cal M}^{(2,2)}_4 & = & \frac{1}{4\rho^4}\left[(3|h_0|^2 - 4|h_1^+|^2 + |h_2^+|^2) + \sqrt{3}(h_0\overline{h_2^+} + \overline{h_0}h_2^+)\right].
\end{eqnarray}
\end{widetext}

\subsection{Trapping probability}

The trapping probability for the five-state Wigner walk can be evaluated using the same method described in Appendix~\ref{app:3state} for the three-state model. The asymptotic value of the amplitude at position $x$ is given by
\begin{equation}
\label{limit:ampl:2}
\psi_\infty^{(2)}(x) = \int\limits_{0}^{2\pi} \frac{dk}{2\pi} e^{- i x k} (v(k),\tilde{\psi}_C) v(k),
\end{equation}
where $\tilde{\psi}_C$ is the Fourier transformation of the initial state
$$
\tilde{\psi}_C = \left(q_2,\ q_{1},\ q_0,\ q_{-1},\ q_{-2}\right)^T,
$$
where the coefficients $q_i$ are given by (\ref{app:sb:2}). The vector $v(k)$ is the eigenstate of the evolution operator in the Fourier picture
$$
\tilde{U}(k) = {\rm Diag}(e^{4i k},\ e^{2i k},\ 1,\ e^{-2i k},\ e^{-4i k})\cdot R^{(2)}(\rho),
$$
corresponding to the eigenvalue 1. Its explicit form is given by
$$
v(k) = \frac{1}{n(k)}\left(
         \begin{array}{c}
           e^{2 i k} \left(1-\rho^2\right) \\
           -\left(1-e^{2 i k}\right) \rho \sqrt{1-\rho^2} \\
           \sqrt{\frac{2}{3}} \left(\rho^2 \cos (2 k)-2 \rho^2+1\right) \\
           \left(1-e^{-2 i k}\right) \rho \sqrt{1-\rho^2} \\
           e^{-2 i k} \left(1-\rho^2\right) \\
         \end{array}
       \right),
$$
where we have denoted by $n(k)$ the normalization factor which reads
$$
n(k) = \sqrt{\frac{2}{3}}(2 - \rho^2 (1 + \cos (2 k))).
$$
One can show by direct calculation that if the initial state is a linear combination only of vectors $|\chi_1^-\rangle$ and $|\chi_2^-\rangle$ then $\tilde{\psi}_C$ is orthogonal to $v(k)$. In such a case, the limiting amplitude vanishes and the trapping effect does not occur. Otherwise, the limiting amplitude (\ref{limit:ampl:2}) can be decomposed into integrals of the form
$$
{\cal I}^{(2)}(x) = \int\limits_{0}^{2\pi} \frac{e^{-i x k}}{4\pi(2-\rho^2(1+\cos{k}))^2} dk,
$$
which is turned into the contour integral over a unit circle in the complex plane and evaluated by the method of residues. We find that the result reads
$$
{\cal I}^{(2)}(x) = Q^{|x|}\frac{2-\rho^2+2|x|\sqrt{1-\rho^2}}{16(1-\rho^2)^\frac{3}{2}},
$$
where $Q$ is given by (\ref{Q}). The trapping probability $p_\infty^{(2)}(x)$ is then obtained by the square norm of the amplitude (\ref{limit:ampl:2}). The final result for positive $x$, negative $x$ and the origin is presented in (\ref{j:2:locp}), (\ref{j:2:locm}) and (\ref{j:2:loc0}).

\end{document}